\newcommand{\be}{\begin{equation}}\newcommand{\ee}{\end{equation}}
\newcommand{\bea}{\begin{eqnarray}}\newcommand{\eea}{\end{eqnarray}}
\newcommand{\nn}{\nonumber\\[6pt]}
\newcommand{\lb}[1]{\label{#1}}
\newcommand{\p}[1]{(\ref{#1})}

\newcommand{\bD}{\overline D}
\newcommand{\cD}{{\cal D}}
\newcommand{\cW}{{\cal W}}
\newcommand{\cV}{{\cal V}}
\newcommand{\cU}{{\cal U}}
\newcommand{\cQ}{{\cal Q}}
\newcommand{\cS}{{\cal S}}
\newcommand{\cT}{{\cal T}}
\newcommand{\cbD}{{\overline{\cal D}}}

\newcommand{\bV}{{\overline V}}
\newcommand{\bQ}{{\overline Q}}
\newcommand{\bS}{{\overline S}}
\newcommand{\bt}{{\bar\theta}}
\newcommand{\bxi}{{\bar\xi}}

\newcommand{\bphi}{{\bar\phi}}
\newcommand{\blam}{{\bar\lambda}}
\newcommand{\bLam}{{\overline\Lambda}}
\newcommand{\bomega}{{\bar\omega}}
\newcommand{\beps}{{\bar\epsilon}}
\newcommand{\vt}{\vartheta}

\documentclass[12pt]{article}
\usepackage{amsmath,amssymb}

\def\theequation{\arabic{section}.\arabic{equation}}
\begin{document}

\thispagestyle{empty}
\vspace{2cm}
\begin{flushright}
hep-th/0312322 \\
ITP-UH-34/03 \\[5mm]
%\today \\[3cm]
\end{flushright}
\begin{center}
{\Large\bf N=8 superconformal mechanics  }
\end{center}
\vspace{1cm}

\begin{center}
{\large\bf S. Bellucci${}^{a}$, E. Ivanov${}^{b}$,
S. Krivonos${}^{b}$, O. Lechtenfeld${}^{c}$ }
\end{center}

\begin{center}
${}^a$ {\it INFN-Laboratori Nazionali di Frascati,
 C.P. 13, 00044 Frascati, Italy}\\
\vspace{0.2cm}

{\tt bellucci@lnf.infn.it}
\vspace{0.4cm}

${}^b$ {\it Bogoliubov  Laboratory of Theoretical Physics, JINR, 141980 Dubna,
Russia}\\
\vspace{0.2cm}

{\tt eivanov, krivonos@thsun1.jinr.ru}

\vspace{0.4cm}
${}^c$ {\it Institut f\"ur Theoretische Physik, Universit\"at Hannover,} \\
{\it Appelstra\ss{}e 2, 30167 Hannover, Germany}
\vspace{0.2cm}

{\tt lechtenf@itp.uni-hannover.de}
\end{center}
\vspace{1cm}

\begin{abstract}
\noindent We construct new models of $N{=}8$ superconformal mechanics
associated with the off-shell $N{=}8, d{=}1$ supermultiplets
$({\bf 3, 8, 5})$ and $({\bf 5, 8, 3})$. These two multiplets are
derived as $N{=}8$ Goldstone superfields and correspond to nonlinear
realizations of the $N{=}8,d{=}1$ superconformal group $OSp(4^\star|4)$
in its supercosets $\frac{OSp(4^\star|4)}{U(1)_R\otimes SO(5)}$ and
$\frac{OSp(4^\star|4)}{SU(2)_R\otimes SO(4)}$, respectively.
The irreducibility constraints for these superfields automatically
follow from appropriate superconformal covariant conditions on the
Cartan superforms. The $N{=}8$ superconformal transformations of the
superspace coordinates and the Goldstone superfields are explicitly given.
Interestingly, each $N{=}8$ supermultiplet admits two different off-shell
$N{=}4$ decompositions, with different $N{=}4$ superconformal subgroups
$SU(1,1|2)$ and $OSp(4^\star|2)$ of $OSp(4^\star|4)$ being manifest as
superconformal symmetries of the corresponding $N{=}4, d{=}1$ superspaces.
We present the actions for all such $N{=}4$ splittings of the $N{=}8$
multiplets considered.
\end{abstract}

\newpage
\setcounter{page}{1}
\section{Introduction}
Supersymmetric quantum mechanics (SQM) \cite{Witt} - \cite{Smi1}
\footnote{See \cite{review} for a more exhaustive list of references.}
has plenty of applications ranging from
the phenomenon of spontaneous breaking of supersymmetry \cite{Witt,ikp,DPRT} to the description of the
moduli of supersymmetric monopoles and black holes \cite{GPS,nscm,mss,BLY}. The latter topic is closely
related to the AdS$_2$/CFT$_1$
pattern of the general AdS/CFT correspondence \cite{ads}, and it is the superconformal versions
of SQM \cite{AP,FR,leva2,nscm,AIPT,GibTo,AKu,IKL,bgk}
which are relevant in this context. Most attention has been paid to SQM models
with extended $N{=}2n$ supersymmetries (where $n{=}1,2,\ldots$) \footnote{By $N$ in SQM models we always
understand the number of {\it real\/} $d{=}1$ supercharges.} because the latter are related to supersymmetries
in $d{>}1$ dimensions via some variant of dimensional reduction.

So far, the mostly explored example is SQM with $N{\leq}4$ supersymmetry. However, the
geometries of $d{=}1$ sigma models having $N{\geq}4$ supercharges are discussed in \cite{GPS,geom}.
Up to now, concrete SQM models with $N{=}8, d{=}1$ supersymmetry have been of particular use.
In \cite{GPS} such sigma models were employed to describe the moduli of certain solitonic black holes.
In \cite{DE}, a superconformal $N{=}8, d{=}1$ action was constructed for the low energy effective dynamics
of a D0-brane moving in D4-brane and/or orientifold plane backgrounds (see also \cite{BMZ,Smi1}).
In \cite{BLY}, $N{=}8$ SQM yielded the low energy description of half-BPS monopoles in $N{=}4$
SYM theory.

It appears to be desirable to put the construction and study of $N{=}8$ (and perhaps $N{>}8$) SQM models on a
systematic basis by working out the appropriate off-shell superfield techniques.

One way to build such models is to perform a direct reduction of four-dimensional superfield theories. For instance, one
may start from a general off-shell action containing $k$ copies of the $d{=}4$ hypermultiplet, naturally written in
analytic harmonic superspace \cite{harm,book}, and reduce it to $d{=}1$ simply by suppressing the dependence
on the spatial coordinates. The resulting action will describe the most general $N{=}8$ extension of a $d{=}1$
sigma model with a $4k$-dimensional hyper-K\"ahler target manifold. Each reduced hypermultiplet yields a $N{=}8, d{=}1$
off-shell multiplet $({\bf 4, 8, \infty })$ \cite{BIK} containing four physical bosons, eight physical fermions and an
infinite number of auxiliary fields just like its $d{=}4$ prototype.

A wider class of $N{=}8$ SQM models can be constructed reducing two-dimensional $N{=}(4,4)$
or heterotic $N{=}(8,0)$
sigma models. In this way one recovers the off-shell $N{=}8, d{=}1$ multiplets $({\bf 4, 8, 4})$
%(from $N{=}(4,4), d{=}2$ twisted chiral multiplet)
and $({\bf 8, 8,0})$ \cite{GR}. The relevant superfield actions describe $N{=}8, d{=}1$ sigma models
with strong torsionful hyper-K\"ahler or octonionic-K\"ahler bosonic target geometries, respectively \cite{GPS}.

Although any $d{=}1$ super Poincar\'e algebra can be obtained from a higher-dimensional
one via dimensional reduction,
this is generally not true for $d{=}1$ super{\it conformal\/} algebras \cite{VP,FRS} and off-shell $d{=}1$ multiplets.
For instance, no $d{=}4$ analog exists for the $N{=}4, d{=}1$ multiplet with off-shell content
$({\bf 1, 4, 3})$ \cite{leva2}
or $({\bf 3, 4, 1})$ \cite{ismi,bepa,mss} (while the multiplet $({\bf 2, 4, 2})$ is actually
a reduction of the chiral $d{=}4$ multiplet).
Moreover, there exist off-shell $d{=}1$ supermultiplets containing no
auxiliary fields at all, something impossible for $d{\geq}3$ supersymmetry. Examples are
the multiplets $({\bf 4,4,0})$ \cite{HP,Hull} and $({\bf 8,8,0})$ \cite{GR} of
$N{=}4$ and $N{=}8$ supersymmetries in $d{=}1$. This zoo of $d{=}1$ supermultiplets greatly enlarges
the class of admissible target geometries for supersymmetric $d{=}1$ sigma models and the related SQM models. Besides
those obtainable by direct reduction from higher dimensions, there exist special geometries particular
to $d{=}1$ models and associated with specific $d{=}1$ supermultiplets.

In fact, dimensional reduction is
not too useful for obtaining super{\it conformally\/} invariant $d{=}1$ superfield actions which are important for
the study of the AdS$_2$/CFT$_1$ correspondence. One of the reasons is that the
integration measures of superspaces having the same Grassmann-odd but different Grassmann-even dimensions possess
different dilatation weights and hence different superconformal transformation properties. As a result,
some superconformal invariant superfield Lagrangians in $d{=}1$ differ in structure from their $d{=}4$
counterparts. Another reason is the already mentioned property that most $d{=}1$ superconformal groups do not descend
from superconformal groups in higher dimensions. The most general $N{=}4, d{=}1$ superconformal group
is the exceptional one-parameter ($\alpha$) family of supergroups $D(2,1;\alpha)$ \cite{FRS} which
only at the special values
$\alpha{=}0$ and $\alpha{=}{-}1$ (and at values equivalent to these two) reduces
to the supergroup $SU(1,1|2)$ obtainable from
superconformal groups $SU(2,2|N)$ in $d{=}4$.

Taking into account these circumstances, it is advantageous to have a convenient superfield
approach to $d{=}1$ models which does not resort to dimensional reduction and is self-contained
in $d{=}1$. Such a framework exists and is based on superfield nonlinear realizations of $d{=}1$ superconformal groups.
It was pioneered in \cite{leva2} and recently advanced in \cite{IKL,IKL2}. Its basic merits are,
firstly, that in most cases it automatically yields the irreducibility conditions for $d{=}1$ superfields
%comprising off-shell $d{=}1$ supermultiplets
and, secondly, that it directly specifies the superconformal
transformation properties of these superfields. The physical bosons and fermions, together with the
$d{=}1$ superspace coordinates,  prove to be coset parameters associated with the appropriate generators of
the superconformal group. Thus, the differences in the field content of various supermultiplets are
attributed to different choices of the coset supermanifold inside the given superconformal group.

Using the nonlinear realizations approach, in \cite{IKL2} all known off-shell multiplets of $N{=}4, d{=}1$ Poincar\'e
supersymmetry were recovered and a few novel ones were found,
including examples of non-trivial off-shell superfield actions.
With the present paper we begin a study of $N{=}8, d{=}1$ supermultiplets along the same line. We will demonstrate that
the $({\bf 5, 8, 3})$ multiplet of ref. \cite{DE} comes out as a Goldstone one, parametrizing a specific coset of
%one of $N{=}8, d{=}1$ superconformal groups, namely
the supergroup $OSp(4^\star|4)$ such that four physical bosons parametrize the coset $SO(5)/SO(4)$ while the fifth one
is the dilaton.  The appropriate irreducibility constraints in $N{=}8, d{=}1$ superspace
immediately follow as a consequence of covariant inverse Higgs \cite{IH} constraints on the relevant Cartan forms.
As an example of
a different $N{=}8$ mechanics, we construct a new model associated with
the $N{=}8, d{=}1$ supermultiplet $({\bf 3, 8, 5})$.
This supermultiplet parametrizes another coset of $OSp(4^\star|4)$ such that $SO(5)\subset OSp(4^\star|4)$ belongs to the
stability subgroup while one out of three physical bosons is the coset parameter associated with the dilatation generator
and the remaining two parametrize the R-symmetry coset $SU(2)_R/U(1)_R$. This model is a direct $N{=}8$ extension
of some particular case of the $N{=}4, d{=}1$
superconformal mechanics considered in \cite{IKL} and the corresponding bosonic sectors are in fact identical.
For both $N{=}8$ supermultiplets considered,
we construct invariant actions in $N{=}4, d{=}1$ superspace. We find an interesting peculiarity of the
$N{=}4, d{=}1$ superfield representation of the model associated with a given $N{=}8$ multiplet.
Depending on the $N=4$ subgroup of the
$N=8$ super Poincar\'e group, with respect to which
we decompose the given $N=8$ superfield, we obtain different splittings
of the latter into irreducible $N{=}4$ off-shell supermultiplets and, consequently,
different $N{=}4$ off-shell actions,
which however produce the same component actions. There exist two distinct $N{=}4$ splittings of the considered
multiplets, namely
\bea
&& ({\bf 5,8,3}) = ({\bf 3,4,1}) \oplus ({\bf 2,4,2}) \qquad\textrm{or}\qquad
   ({\bf 5,8,3}) = ({\bf 1,4,3}) \oplus ({\bf 4,4,0}) \ ,\quad\label{1} \\[8pt]
&& ({\bf 3,8,5}) = ({\bf 1,4,3}) \oplus ({\bf 2,4,2}) \qquad\textrm{or}\qquad
   ({\bf 3,8,5}) = ({\bf 3,4,1}) \oplus ({\bf 0,4,4}) \ .\quad\label{2}
\eea
The first splitting in \p{1} is just what has been employed in \cite{DE}.
The second splitting is new. For both of them we write down $N{=}8$ superconformal actions
in standard $N{=}4, d{=}1$ superspace.
The latter is also suited for setting up the off-shell superconformal action corresponding to the first option in
\p{2}. As for the second splitting in \p{2}, the equivalent off-shell superconformal action can be written only
by employing $N{=}4, d{=}1$ harmonic superspace \cite{IL}, since the kinetic term of the multiplet
$({\bf 0, 4, 4})$ is naturally defined just in this superspace.

The paper is organized as follows. In Section 2 we give a $N{=}8$ superfield formulation of the multiplet
$({\bf 3, 8, 5})$, based on a nonlinear realization of the superconformal supergroup $OSp(4^\star|4)$.
In Sections 3 and 4 we
present two alternative $N{=}4$ superfield formulations of this multiplet and the relevant
off-shell superconformal actions.
Section 5 is devoted to treating the multiplet $({\bf 5, 8, 3})$ along the same lines. A summary of our results
and an outlook are the contents of the concluding Section 6.

\section{N=8, d=1 superspace and tensor multiplet}

\subsection{N=8, d=1 superspace as a reduction of N=2, d=4}
It will be more convenient for us to start from the $N=8, d=1$ superfield description of the off-shell
multiplet $({\bf 3, 8, 5})$, since it is tightly related to the $N=4, d=1$ model studied in \cite{IKL}. We
shall first recover it within the dimensional reduction from $N=2, d=4$ superspace.

The maximal automorphism group of $N=8, d=1$ super Poincar\'e algebra (without central charges) is $SO(8)$
and so eight real Grassmann coordinates of $N=8, d=1$ superspace can be arranged into one of three 8-dimensional
real irreps of $SO(8)$.
For our purpose in this paper we shall split these 8 coordinates in another way, namely into two real quartets
on which three commuting automorphism $SU(2)$ groups will be realized.

A convenient point of departure is $N=2, d=4$ superspace
$(x^m, \theta_i^\alpha, \bar\theta^{i\dot\alpha})$ whose automorphism group is $SL(2,C)\times U(2)_R$. The corresponding
covariant derivatives form the following algebra: \footnote{We use the following convention for the skew-symmetric
tensor $\epsilon$: $\;\epsilon_{ij} \epsilon^{jk}=\delta_i^k \;, \quad \epsilon_{12} = \epsilon^{21} =1 \;$.}
\be\label{t2}
\left\{ \cD^{i}_{\alpha},\cbD^{j}_{\dot\alpha}\right\} =2i \epsilon^{ij}\partial_{\alpha\dot\alpha}\,,
\ee
where $\cbD_{j\dot\alpha} \equiv - \overline{(\cD^j_\alpha)}$. The $N=2, d=4$ tensor multiplet is described
by a real isotriplet superfield $\cV^{(ik)}$ subjected to the off-shell constraints
\be\label{t1}
\cD^{(i}_\alpha \cV^{jk)}=\cbD^{(i}_{\dot\alpha} \cV^{jk)}=0\,.
\ee

These constraints can be reduced to $d=1$ in two different ways, yielding multiplets of two different
$d=1$ supersymmetries.

To start with, it has been recently realized \cite{IKL2} that many $N=4, d=1$ superconformal multiplets have
their $N=2, d=4$ ancestors (though the standard dimensional reduction to $N=4, d=1$ superspace proceeds from
$N=1, d=4$ superspace, see e.g. \cite{ismi}). The relation between such $d=4$ and $d=1$ supermultiplets is
provided by a
special reduction procedure, whose key feature is the suppression of space-time indices in $d=4$
spinor derivatives. The superfield constraints describing the $N=4, d=1$ supermultiplets are obtained from
the $N=2, d=4$ ones by discarding the spinorial $SL(2,C)$ indices and keeping only the R-symmetry indices.
For instance, the constraints \p{t1} after such a reduction become
\be\label{t3}
D^{(i}V^{jk)}=\bD^{(i} V^{jk)}=0\,.
\ee
Here, the `reduced' $N=4, d=1$ spinor derivatives $D^i, \bD^j$ are subject to
\be\label{t4}
\left\{ D^i, \bD^j\right\} = 2i\epsilon^{ij} \partial_t
\ee
and $\partial_{\alpha\dot\alpha} \rightarrow \partial_t$. The $N=4, d=1$ superfield $V^{ij}$
obeying the constraints \p{t3} contains four bosonic (three
physical and one auxiliary) and four fermionic off-shell components, i.e. it defines the $N=4, d=1$
multiplet $({\bf 3,4,1})$. It has been employed in \cite{ismi} for constructing a general off-shell sigma
model corresponding to the $N=4$ SQM model of \cite{CR} and
\cite{Smi0}. The same supermultiplet was independently considered in \cite{bepa,mss} and then has been used
in \cite{IKL} to construct a new version of $N=4$ superconformal mechanics. It has been also treated in
the framework of $N=4, d=1$ harmonic superspace \cite{IL}.

Other $N=2, d=4$ supermultiplets can be also reduced in this way to yield their $N=4, d=1$ superspace
analogs \cite{IKL2}. It should be emphasized that, though formally reduced constraints look similar to their
$N=2, d=4$ ancestors, the irreducible component field contents of the relevant multiplets can radically differ
from those in $d=4$ due to the different structure of the algebra of the covariant derivatives. For example, \p{t1}
gives rise to a notoph type differential constraint for a vector component field, while \p{t3} does not impose any
restriction on the $t$-dependence of the corresponding component fields. As an extreme expression of such a relaxation
of constraints, some $N=2, d=4$ supermultiplets which are  on-shell in the standard $d=4$ superspace in consequence of
their superfield constraints, have off-shell $N=4, d=1$ counterparts. This refers e.g. to the $N=2, d=1$ hypermultiplet
without central charge \cite{Sohn} (leaving aside the formulations in $N=2, d=4$ harmonic superspace \cite{harm,book}).

Another, more direct way to reduce some constrained $N=2, d=4$ superfield to $d=1$ is to keep as well
the $SL(2,C)$ indices of spinor derivatives, thus preserving the total number of spinor coordinates and
supersymmetries and yielding a $N=8, d=1$ supermultiplet. However, such a reduction clearly breaks $SL(2,C)$ down
to its $SU(2)$ subgroup
\be\label{T2}
\left\{ \cD^{i\alpha},\cbD^{j\beta}\right\} =2i \epsilon^{ij}\epsilon^{\alpha\beta}\partial_t\,.
\ee
A closer inspection of the relation \p{T2} shows that, besides the $U(2)_R$ automorphisms \footnote{Hereafter,
we reserve the term `R-symmetry' for those automorphisms of $d=1$ Poincar\'e superalgebra which originate
from R-symmetries in $d=4$.} and
the manifest $SU(2)$ realized on the indices $\alpha, \beta$ and inherited from $SL(2,C)$,
it also possesses the hidden automorphisms
\be
\delta \cD^i_\alpha = \Lambda_\alpha^\beta \cbD^i_\beta, \quad \delta \cbD^i_\alpha =
-\bar{\Lambda}_\alpha^\beta \cD^i_\beta, \quad
\Lambda^\alpha_\alpha = 0\,, \label{T3}
\ee
which emerge as a by-product of the reduction. Together with the manifest $SU(2)$ and
overall phase $U(1)_R$ transformations, \p{T3} can be shown
to form $USp(4) \sim SO(5)$. The transformations with
$\bar{\Lambda}^\alpha_\beta = - \Lambda^\alpha_\beta$ close on the manifest
$SU(2)$ and, together with the latter, constitute the subgroup
$Spin(4) = SU(2)\times SU(2) \subset USp(4)$. The manifest $SU(2)$ forms
a diagonal in this product. Passing to the new basis
\be
D^{ia} = {1\over \sqrt{2}}\left( \cD^{ia} + \cbD{}^{ia} \right), \quad \nabla^{i\alpha} =
{i\over \sqrt{2}}\left( \cD^{i\alpha} - \cbD{}^{i\alpha} \right)\,, \overline{(D^{ia})} =
-D_{ia}\,,\;\;\overline{(\nabla^{i\alpha})} =
-\nabla_{i\alpha}\,, \label{Pass}
\ee
one can split \p{T2} into two copies of $N=4, d=1$ anticommutation relations, such that the covariant derivatives
from these sets anticommute with each other
\be\label{m1}
\left\{ D^{ia},  D^{jb} \right\} = 2i \epsilon^{ij}\epsilon^{ab} \partial_t\,,\quad
\left\{ \nabla^{i\alpha}, \nabla^{j\beta} \right\} = 2i \epsilon^{ij}\epsilon^{\alpha\beta} \partial_t\,, \quad
\left\{D^{ia}, \nabla^{j\alpha} \right\}=0\,.
\ee
These two mutually anticommuting algebras pick up the left and right $SU(2)$ factors of $Spin(4)$
as their automorphism symmetries. Correspondingly, the $N=8, d=1$ superspace is parametrized by the
coordinates $(t, \theta_{ia}, \vartheta_{k\alpha})$, subjected to the reality conditions
\be
\overline{(\theta_{ia})} = \theta^{ia}\,, \quad \overline{(\vartheta_{i\alpha})} = \vartheta^{i\alpha}\,.
\ee
Such a representation of the algebra of $N=8, d=1$ spinor derivatives manifests three mutually commuting automorphism
$SU(2)$ symmetries which are realized, respectively, on the doublet indices $i, a$ and $\alpha$. The transformations from
the coset $USp(4)/Spin(4) \sim SO(5)/SO(4)$ rotate $D^i_a$ and $\nabla^i_\alpha$ through each other (and the same
for $\theta_{ia}$ and $\vartheta_{i\alpha}$).

In this basis, the $N=8, d=1$ reduced version of $N=2, d=4$ tensor multiplet \p{t1} is defined by
the constraints
\be\label{m2}
D^{(i}_a V^{jk)} = \nabla^{(i}_\alpha V^{jk)}=0 \;.
\ee
This off-shell $N=8, d=1$ multiplet can be shown to comprise eight bosonic (three physical and five auxiliary)
and eight fermionic components, i.e. it is $({\bf 3, 8, 5})$. We shall see that \p{m2} implies a $d=1$ version
of the $d=4$ notoph field strength constraint whose effect in $d=1$ is to constrain some superfield component
to be constant \cite{leva2}.

In the next Section we shall show that the $N=8, d=1$ tensor multiplet $V^{ij}$ defined by \p{m2} can support,
besides the manifest $N=8, d=1$ Poincar\'e supersymmetry, also a realization of the $N=8, d=1$ superconformal
algebra $osp(4^\star |4)$. While considered as a carrier of the latter, it can be called
`$N=8, d=1$ improved tensor multiplet'.
In fact, like its $N=4, d=1$ counterpart $({\bf 3, 4, 1})$, it can
be derived from a nonlinear realization of the supergroup $OSp(4^\star |4)$ in the appropriate coset
supermanifold, without any reference to the dimensional reduction from $d=4$.

\subsection{Superconformal properties of the N=8, d=1 tensor multiplet}

The simplest way to find the transformation properties of the $N=8, d=1$ tensor multiplet
$V^{ij}$ and prove the covariance of the basic constraints \p{m2} with respect to the
superconformal $N=8$ superalgebra  $osp(4^\star |4)$ is to use the coset realization technique.
All steps in this construction are very similar to those employed in \cite{IKL}. So
we quote here the main results without detailed explanations.

We use the standard definition  of the superalgebra $osp(4^\star |4)$ \cite{FRS}.
It contains the following sixteen spinor generators:
\be\label{alg1}
Q_1^{iaA},\; Q_2^{i\alpha A},\quad \overline{\left( Q^{iaA}\right)}=\epsilon_{ij}\epsilon_{ab}Q^{jbA},
\quad (i,a,\alpha, A=1,2),
\ee
and sixteen bosonic generators:
\be\label{alg2}
T_0^{AB},\quad T^{ij},\quad T_1^{ab},\quad T_2^{\alpha\beta},\quad U^{a\alpha}\;.
\ee
The indices $A,i,a$ and $\alpha$ refer to fundamental representations of the mutually commuting
$sl(2,R)\sim  T_0^{AB}$ and three $su(2)\sim T^{ij}, T_1^{ab},T_2^{\alpha\beta} $ algebras.
The four generators $U^{a\alpha}$ belong to the coset $SO(5)/SO(4)$ with $SO(4)$ generated by
$T_1^{ab}$ and $T_2^{\alpha\beta}\,$. The  bosonic generators form
the full bosonic subalgebra $sl(2,R)\oplus su(2)_R\oplus so(5)$ of $osp(4^\star |4)$.

The commutator of any $T$-generator with $Q$ has the same form, and it is sufficient to
write it for some particular sort of indices, e.g. for $a,b$ (other indices of
$Q_{1,2}^{iaA}$ being suppressed):
\be\label{alg5}
\left[ T^{ab}, Q^{c}\right]= -\frac{i}{2}\left( \epsilon^{ac}Q^{b}+
 \epsilon^{bc}Q^{a} \right).
\ee
The commutators with the coset $SO(5)/SO(4)$ generators $U^{a\alpha}$ have
the following form
\be\label{alg6}
\left[ U^{a\alpha}, Q_1^{ibA}\right]=-i\epsilon^{ab}Q_2^{i\alpha A},\quad
\left[ U^{a\alpha}, Q_2^{i\beta A}\right]=-i\epsilon^{\alpha\beta}Q_1^{iaA}\;.
\ee
At last, the anticommutators of the fermionic generators read~
\bea\label{alg7}
&&\left\{ Q_1^{iaA}, Q_1^{jbB}\right\} = -2\left(
  \epsilon^{ij}\epsilon^{ab}T_0^{AB}  -2
  \epsilon^{ij}\epsilon^{AB}T_1^{ab}+
  \epsilon^{ab}\epsilon^{AB}T^{ij}
\right), \nn
&&\left\{ Q_2^{i\alpha A}, Q_2^{j\beta B}\right\} = -2\left(
  \epsilon^{ij}\epsilon^{\alpha\beta}T_0^{AB}  -2
  \epsilon^{ij}\epsilon^{AB}T_2^{\alpha\beta}+
  \epsilon^{\alpha\beta}\epsilon^{AB}T^{ij}
\right), \nn
&&\left\{ Q_1^{iaA}, Q_2^{j\alpha B}\right\} =2\epsilon^{ij}\epsilon^{AB}U^{a\alpha}\;.
\eea
{}From \p{alg7} it follows that the generators $Q_1^{iaA}$ and $Q_1^{i\alpha A}$, together with the corresponding
bosonic generators, span two $osp(4^\star |2)$ subalgebras in $osp(4^\star |4)$.
For what follows it is convenient to pass to another notation,
\bea
&& P\equiv T_0^{22}, \; K \equiv  T_0^{11}, \; D\equiv -T_0^{12}, \quad
 V\equiv T^{22}, \; \bV\equiv  T^{11},\; V_3 \equiv  T^{12}, \nn
&& Q^{ia} \equiv -Q_1^{ia2},\; \cQ^{i\alpha}\equiv -Q_2^{i\alpha 2},\; S^{ia}\equiv Q_1^{ia1},\;
\cS^{i\alpha}\equiv Q_2^{i\alpha 1}\,. \label{alg8}
\eea
One can check that $P$ and $Q^{ia}, \cQ^{i\alpha}$ constitute a $N=8, d=1$ Poincar\'e superalgebra.
The generators $D, K$ and  $S^{ia}, \cS^{i\alpha}$ stand for the $d=1$ dilatations, special conformal
transformations and conformal supersymmetry, respectively. The full structure of the superalgebra $osp(4^\star|4)$
is given in Appendix.

Now we shall construct a nonlinear realization  of the superconformal group $OSp(4^\star |4)$
in the coset superspace with an element parametrized as
\be\label{coset}
g=e^{itP}e^{\theta_{ia} Q^{ia}+\vt_{i\alpha} \cQ^{i\alpha}}e^{\psi_{ia} S^{ia}+\xi_{i\alpha} \cS^{i\alpha}}
e^{izK}e^{iuD}e^{i\phi V+ i\bphi \bV}~.
\ee
The coordinates $t, \theta_{ia}, \vt_{i\alpha}$ parametrize the $N=8, d=1$ superspace. All other
supercoset parameters are Goldstone $N=8$ superfields. The stability subgroup
contains a subgroup $U(1)_R$ of the group $SU(2)_R$ realized on the doublet indices $i$, hence
the Goldstone superfields $\phi, \bar\phi$ parametrize the coset $SU(2)_R/U(1)_R$. The
group $SO(5)$ is placed in the stability subgroup. It linearly rotates the fermionic
Goldstone superfields $\psi$ and $\xi$ through each other, equally as the
$N=8, d=1$ Grassmann coordinates $\theta$ and $\vt$'s. To summarize, in the present case we are dealing with
the supercoset $\frac{OSp(4^\star|4)}{U(1)_R\otimes SO(5)}$.

The semi-covariant (fully covariant only under $N=8$ Poincar\'e supersymmetry)
spinor derivatives are defined by
\be
D^{ia}=\frac{\partial}{\partial\theta_{ia}}+i\theta^{ia} \partial_t\; , \;
\nabla^{i\alpha}=\frac{\partial}{\partial\vt_{i\alpha}}+i\vt^{i\alpha} \partial_t\;.
\ee
Their anticommutators, by construction, coincide with \p{m1}.

A natural way to find conformally covariant irreducibility conditions on the coset superfields
is to impose the inverse Higgs constraints \cite{IH} on
the left-covariant Cartan one-form $\Omega$ valued in the superalgebra $osp(4^\star |4)$.
This form is defined by the standard relation
\be
g^{-1}\,d\,g = \Omega~.
\ee
In analogy with \cite{IKL,IKL2}, we impose the following constraints:
\be\label{ih}
\omega_D=0\; , \quad \left. \left . \omega_V\right| = \bomega_V\right|=0
\ee
where $|$ denotes the spinor projection. These constraints are manifestly covariant
under the whole supergroup. They allow one to express the Goldstone spinor superfields
and the superfield $z$ via the spinor and $t$-derivatives, respectively,
of the remaining bosonic Goldstone superfields $u, \phi, \bar\phi\,$
\bea\label{mc}
&& i D^{ia} u = -2 \psi^{ia}\; , \; i\nabla^{i\alpha} u=-2 \xi^{i\alpha} \; , \; \dot{u}=2z\; ,\nn
&& i D^{1a} \Lambda = 2\Lambda\left( \psi^{1a} + \Lambda \psi^{2a}\right), \;
 i D^{2a} \Lambda = -2\left( \psi^{1a} + \Lambda \psi^{2a}\right),  \nn
&& i \nabla^{1\alpha} \Lambda = 2\alpha\left( \xi^{1\alpha} + \Lambda \xi^{2\alpha}\right), \;
 i \nabla^{2\alpha} \Lambda = -2\Lambda\left( \xi^{1\alpha} + \Lambda \xi^{2\alpha}\right),
\eea
where
\be \label{Lambda}
\Lambda = \frac{ \tan \sqrt{\phi\bphi}}{\sqrt{\phi\bphi}}\phi \; ,\;
\bLam = \frac{ \tan \sqrt{\phi\bphi}}{\sqrt{\phi\bphi}}\bphi \; .
\ee
Simultaneously, eqs. \p{mc} imply some irreducibility constraints for $u, \phi, \bar\phi\,$.
After introducing a new $N=8$ vector superfield $V^{ij}$ such that
$V^{ij}=V^{ji}$ and $\overline{V^{ik}}=\epsilon_{ii'}\epsilon_{kk'}V^{i'k'}$, via
\bea\label{V}
&& V^{11}=-i\sqrt{2}\,e^{u}\frac{\Lambda}{1+\Lambda\bLam} \; ,\;
V^{22}= i\sqrt{2}\,e^{u}\frac{\bLam}{1+\Lambda\bLam}\; , \;
V^{12}=  \frac{i}{\sqrt{2}}\,e^{u}\frac{1-\Lambda\bLam}{1+\Lambda\bLam}\; , \nonumber \\
&& V^2 \equiv V^{ik}V_{ik} = e^{2u}~,
\eea
and eliminating the spinor superfields from \p{mc}, the differential constraints on the remaining
Goldstone superfields can be brought in the manifestly $SU(2)_R$-symmetric form
\be\label{tensor}
D_a^{(i}V^{jk)} =0 \; , \quad \nabla_\alpha{}^{(i}V^{jk)} =0 \; ,
\ee
which coincides with \p{m2}. For further use, we present one important consequence of \p{tensor}
\be
\partial_t\left(D^a_i D_{ja}V^{ij} + \nabla^\alpha_i \nabla_{j\alpha}V^{ij}\right) = 0 \;\Rightarrow \;
\nabla^\alpha_i \nabla_{j\alpha}V^{ij} = 6m - D^a_i D_{ja}V^{ij}\,, \; m = \mbox{const}. \lb{constm}
\ee
The specific normalization of the arbitrary real constant $m$ is chosen for convenience.

Besides ensuring the covariance of the basic constraints \p{tensor} with respect to
$OSp(4^\star |4)$ and clarifying their geometric meaning, the coset approach provides the easiest
way to find the transformation properties of all coordinates and superfields. Indeed, all
transformations are generated by acting on the coset element \p{coset} from the left by the
elements of $osp(4^\star |4)$. Since all bosonic transformations appear in the anticommutator of the conformal
supersymmetry and Poincar\'e supersymmetry, it is sufficient to know how $V^{ij}$ is transformed under these
supersymmetries.

The $N=8, d=1$ Poincar\'e supersymmetry is realized on superspace coordinates in the standard way
\be\label{n8p}
\delta t = -i\left( \eta_{ia}\theta^{ia}+\eta_{i\alpha}\vt^{i\alpha}\right) ,
 \quad \delta \theta_{ia}=\eta_{ia},\quad \delta\vt_{i\alpha}=\eta_{i\alpha}
\ee
and $V^{ik}$ is a scalar with respect to these transformations. On the other hand,
$N=8$ superconformal transformations are non-trivially realized both on the coordinates and coset superfields
\bea\label{n8s}
&& \delta t = -it\left(\epsilon^{ia}\theta_{ia}+\varepsilon^{i\alpha}\vt_{i\alpha}\right)
+\left(\epsilon^i_a \theta^{ja}+\varepsilon^i_\alpha \vt^{j\alpha}\right)\left(
\theta_{ib}\theta^b_j +\vt_{i\beta}\vt^\beta_j\right)\;, \nn
&& \delta \theta_{ia}=t \epsilon_{ia}-i\epsilon^j_a\theta_{jb}\theta^b_i+
2i \epsilon^j_b \theta^b_i\theta_{ja}-i\epsilon^j_a \vt_{j\alpha}\vt^\alpha_i
+2i \varepsilon^\alpha_j\vt_{i\alpha}\theta^j_a \; , \nn
&& \delta\vt_{i\alpha}=t\varepsilon_{i\alpha}-i\varepsilon^j_\alpha \vt_{j\beta}\vt^\beta_i
+2i \varepsilon^j_\beta \vt^\beta_i\vt_{j\alpha}-i\varepsilon^j_\alpha \theta_{ja}\theta^a_i
+2i \epsilon^j_a \theta^a_i\vt_{j\alpha} \; , \nn
&& \delta u = -2i\left( \epsilon^{ia}\theta_{ia}+\varepsilon^{i\alpha} \vt_{i\alpha}\right)\; , \quad
\delta \Lambda = a+ib\Lambda +{\bar a} \Lambda^2\;,
\eea
where
\bea
 && a=2i\left( \epsilon^a_2\theta_{2a}+\varepsilon^\alpha_2\vt_{2\alpha}\right),\;
{\bar a}=2i\left( \epsilon^a_1\theta_{1a}+\varepsilon^\alpha_1\vt_{1\alpha}\right),\nn
&& b=-2\left(\epsilon^a_1\theta_{2a}+\epsilon^a_2\theta_{1a}+\varepsilon^\alpha_1\vt_{2\alpha}+
\varepsilon^\alpha_2 \vt_{1\alpha} \right)\;.
\eea
The conformal supersymmetry transformation of $V^{ij}$ has the manifestly $Spin(4)\times SU(2)_R$ covariant form
\be\label{Vtr}
\delta V^{ij}=2i\left[(\theta^{ka}\epsilon_{ka}+\vt^{k\alpha}\varepsilon_{k\alpha})V^{ij}+
(\epsilon^{a(i}\theta_{ka}+\epsilon^a_k\theta^{(i}_a +
\varepsilon^{\alpha (i}\vt_{k\alpha}+\varepsilon^{\alpha}_k \vt^{(i}_\alpha )V^{j)k}\right].
\ee

Thus we know the transformation properties of the $N=8$ `tensor' multiplet $V^{ij}$ under
the $N=8$ superconformal group $OSp(4^\star |4)$. Before turning to the construction of superconformal
invariant actions, in the next Section we will study how this $N=8$ multiplet is described
in $N=4, d=1$ superspace.

\setcounter{equation}0
\section{N=8, d=1 tensor multiplet in N=4 superspace}

While being natural and very useful for deriving irreducibility constraints for the superfields
and establishing their transformation properties, the $N=8, d=1$ superfield
approach is not too suitable for constructing invariant actions. The basic difficulty is, of course,
the large dimension of the integration measure, which makes it impossible to write the action in terms
of this basic superfields without introducing the prepotential and/or passing to some invariant superspaces
of lower Grassmann dimension, e.g. chiral or harmonic analytic superspaces.
Another possibility is to formulate the $N=8$ tensor multiplet in $N=4, d=1$ superspace. In such a formulation
half of $N=8$ supersymmetries is hidden and only $N=4$ supersymmetry is manifest. Nevertheless, it allows
a rather straightforward construction of $N=8$ supersymmetric and superconformal actions. Just this approach
was used in \cite{DE}. In the next sections we shall reproduce the action of \cite{DE} and construct new $N=4$ superfield
actions with second hidden $N=4$ supersymmetry, both for vector and tensor $N=8$ multiplets. In the present section
we consider two $N=4$ superfield formulations of the $N=8$ tensor multiplet.

\subsection{(3,8,5) = (3,4,1)$\oplus$(0,4,4)}
In order to describe the $N=8$ tensor multiplet in terms of $N=4$ superfields we should choose the
appropriate $N=4$ superspace. The first (evident) possibility is to consider the $N=4$ superspace
with coordinates
\be
\left( t, \theta_{ia} \right). \lb{Isup}
\ee
In this superspace the $N=4$ conformal supergroup
$$
OSp(4^\star |2)\sim \left\{ P,K,D, T^{ij}, T_1^{ab}, Q^{ia}, S^{ia}\right\}
$$
is naturally realized, while the rest of the $osp(4^\star |4)$ generators mixes two irreducible $N=4$
superfields comprising the $({\bf 3, 8, 5})$ $N=8$ supermultiplet in question. Expanding the $N=8$ superfields $V^{ij}$
in $\vt_{i\alpha}$,  one finds that the constraints \p{tensor} leave in $V^{ij}$ the following
four bosonic and four fermionic $N=4$ projections:
\be\label{n4a}
\left. v^{ij}=V^{ij}\right| ,\quad \left. \xi^i_\alpha \equiv \nabla_{j\alpha}V^{ij}\right|,\quad
\left. A\equiv \nabla^\alpha_i \nabla_{j\alpha}V^{ij}\right|
\ee
where $|$ means restriction to $\vt_{i\alpha}=0$. Each $N=4$ superfield is subjected, in virtue
of \p{tensor}, to an additional constraint
\bea
&& D_a^{(i}v^{jk)}=0\,,\quad  D_a^{(i}\xi^{j)}_\alpha =0\,,\nn
&& A= 6m -D^a_i D_{aj} v^{ij}\,, \;m = \mbox{const}\,, \lb{n4b}
\eea
where we used \p{constm}.

Thus, we conclude that our $N=8$ tensor multiplet $V^{ij}$, when rewritten in terms of $N=4$ superfields,
amounts to a direct sum of the $N=4$ `tensor' multiplet $v^{ij}$ with the $({\bf 3,4,1})$ off-shell content and a fermionic analog
of the $N=4$ hypermultiplet $\xi^i_\alpha$ with the $({\bf 0,4,4})$ off-shell content,\footnote{This
supermultiplet has been introduced in \cite{GR}; its off-shell superfield action was given in \cite{IL} in the framework of $N=4, d=1$
harmonic superspace.} plus a constant
$m$ of dimension of mass (i.e. $(cm^{-1})$). The conservation-law type condition \p{constm} yielding this constant is in fact
a $d=1$ analog of the `notoph' condition $\partial_mA^m(x) = 0$ of the $N=2, d=4$ tensor multiplet; the appearance of a similar
constant in the case of the $N=4, d=1$ supermultiplet $({\bf 1, 4, 3})$, which is defined by the $d=1$ reduction
of the $N=1, d=4$ tensor multiplet constraints, was earlier observed in \cite{leva2}.

The transformations of the implicit $N=4$ Poincar\'e supersymmetry completing the manifest
one to the full $N=8$ Poincar\'e supersymmetry have the following form in terms of $N=4$ superfields:
\be\label{n4transf}
\delta^* v^{ij}=\eta^{(i}\xi^{j)}-\overline{\eta}{}^{(i}\bxi^{j)}\,,\quad
\delta^*\xi^i=-2i\overline{\eta}_j{\dot v}{}^{ij}-\frac{1}{3}\overline{\eta}{}^i
D_j\bD_kv^{jk}+6m\,\overline{\eta}^i\;,
\ee
where we passed to standard $N=4, d=1$ derivatives \p{t4}
\be
D^i \equiv D^{i1},\quad \bD^i \equiv D^{i2} \lb{defD1}
\ee
and the complex notation
\be
\eta^i \equiv \eta^{1i}\,, \;\; \overline{\eta}^i \equiv \eta^{2i}\,,\quad \lb{complnot}
\xi^i \equiv \xi^{1i}\,,\;\; \bxi^i \equiv \xi^{2i} \,.
\ee
Since any other $osp(4^\star|4)$ transformation appears in the anticommutator of $N=4$ conformal supersymmetry
(realized as in \p{n8s}-\p{Vtr} with $\varepsilon=\vt=0$) and implicit $N=4$  Poincar\'e supersymmetry
transformations, it suffices to require invariance under these two basic supersymmetries when constructing
invariant actions for the considered system in the $N=4, d=1$ superspace. Note that we could equally choose
the $N=4, d=1$ superspace $(t, \vartheta_{i\alpha})$ to deal with, and expand $V^{ij}$ with respect to
$\theta_{ia}$. The final $N=4, d=1$ splitting of the multiplet $({\bf 3,8,5})$ will be the same modulo
the interchange $\alpha \;\leftrightarrow a$. These two `mirror' choices manifestly preserve covariance under the
R-symmetry group $SU(2)_R$ realized on the indices $i, k$.

\subsection{(3,8,5) = (1,4,3)$\oplus$(2,4,2)}
Surprisingly, there is a more sophisticated choice of a $N=4$ subspace in the $N=8, d=1$ superspace which
gives rise to a different $N=4$ splitting of the considered $N=8$ supermultiplet, namely, that into the
multiplets $({\bf 1, 4, 3})$ and $({\bf 2, 4, 2})$.  It corresponds to dividing
the $N=8, d=1$ Grassmann coordinates with respect to the $SU(2)_R$ doublet indices. As a result, the $SU(2)_R$
symmetry becomes implicit, as opposed to the previous options.

The essential difference between two possible $N=4$ splittings is closely related to the fact that
they are covariant under different $N=4, d=1$ superconformal groups. The most general $N=4, d=1$ superconformal
group is known to be $D(2,1;\alpha)$ and its (anti)commutation relations involve an arbitrary parameter $\alpha$.
The $N=4, d=1$ superconformal group $OSp(4^\star|2) \subset OSp(4^\star|4)$ which is explicit in the $N=4, d=1$
superfield formulation discussed in the previous subsection is just $D(2,1;\alpha=1)\sim OSp(4^\star|2)$.
One can wonder whether
$OSp(4^\star|4)$ contains as a subgroup also some other special case of $D(2,1;\alpha)$
corresponding to a different choice of $\alpha$.
Our approach immediately implies that $OSp(4^\star|4)$ contains another $N=4, d=1$ superconformal
group $SU(1,1|2)\sim D(2,1;\alpha=-1)$. Although this fact can be of course established by an inspection of
the root diagrams for $OSp(4^\star|4)$, in our approach it is visualized in the possibility of the just mentioned
alternative splitting of the multiplet $({\bf 3, 8, 5})$ into the sum of the off-shell $N=4, d=1$ multiplets
$({\bf 1, 4, 3})$ and $({\bf 2, 4, 2})$. The latter is a chiral $N=4$ multiplet, while the only $N=4, d=1$
superconformal group which respects chirality is just $SU(1,1|2)$ \cite{IKL,IL}.

Using the (anti)commutation relations \p{alg5}-\p{alg7} one can check that the following
fermionic generators:
\be\label{su1}
Q^a=Q_1^{1a2}+iQ_2^{1a2},\;\bQ^a=Q_1^{1a2}-iQ_2^{1a2},\; S^a=Q_1^{1a1}+iQ_2^{1a1},\;
\bS^a=Q_1^{1a1}-iQ_2^{1a1}\;,
\ee
together with the bosonic ones
\be\label{su2}
P,\;D,\;K,\;{\cal T}^{ab}\equiv T_1^{ab}+T_2^{ab},\; Z\equiv U^{12}-U^{21}-2iT^{12}\;,
\ee
indeed form a $su(1,1|2)$ superalgebra with central charge $Z$.
Let us note that the only $su(2)$ subalgebra which remains manifest in this basis
is the diagonal $su(2)\sim \cT^{ab}$ in $o(4)\sim \left( T_1^{ab},T_2^{\alpha\beta}\right)$.

The splitting of the $N=8$ superfield $V^{ij}$ in terms of the $N=4$ ones in the newly introduced basis can be
performed  as follows. Firstly, we define the new covariant derivatives with the indices of the fundamental
representations of the diagonal $su(2) \subset Spin(4)$ as
\bea\label{sucovder}
&&  D^a\equiv \frac{1}{\sqrt{2}}\left( D^{1a}+i\nabla^{1a}\right),\;
\bD_a\equiv \frac{1}{\sqrt{2}}\left( D_a^{2}-i\nabla_a^{2}\right), \nn
&& \nabla^a\equiv \frac{i}{\sqrt{2}}\left( D^{2a}+i\nabla^{2a}\right),\;
\bar\nabla_a\equiv \frac{i}{\sqrt{2}}\left( D_a^{1}-i\nabla_a^{1}\right),
\eea
with the only non-zero anticommutators being
\be\label{sucovder1}
\left\{D^a,\bD_b \right\}=-2i\delta^a_b\partial_t\,,\quad
\left\{\nabla^a,\bar\nabla_b \right\}=-2i\delta^a_b\partial_t\, .
\ee

Now, as an alternative $N=4$ superspace we choose the set of coordinates
closed under the action of $D^a, \bar D_a$, i.e.
\be
\left( t\,,\; \theta_{1a} - i\vartheta_{1a}\,,\; \theta^{1a} + i \vartheta^{1a} \right), \lb{IIsup}
\ee
while the $N=8$ superfields are expanded with respect to the orthogonal combinations $\theta^a_2 - i \vartheta^a_2\,$,
$\theta^a_1 + i\vartheta^a_1$ annihilated by $D^a, \bar D_a$.

The basic constraints \p{tensor}, being rewritten through \p{sucovder}, read
\bea\label{sumultiplet}
&& D^a\varphi=0\,, \quad D^a v -\nabla^a \varphi=0\,,
\quad \nabla^a v + D^a{\bar\varphi}=0\,,\quad \nabla^a \bar\varphi =0\,, \nn
&& {\overline{\nabla}}_a \varphi=0,\quad
\overline{\nabla}_a v+\bD_a \varphi=0\,,\quad \bD_a v - \overline{\nabla}_a \bar\varphi =0\,,
\quad \bD_a\bar\varphi=0\;,
\eea
where
\be\label{v}
v\equiv -2i V^{12}\,,\quad \varphi \equiv V^{11}\,,\quad \bar\varphi\equiv V^{22}\,.
\ee
Due to the constraints \p{sumultiplet}, the derivatives $\nabla^a$ and $\overline{\nabla}_a$ of every $N=8$
superfield in the triplet $\left(V^{12}, V^{11}, V^{22} \right)$ can be expressed as $D^a, \bD_a$
of some other superfield. Therefore, only
the zeroth order (i.e. taken at $\theta^a_2-i\vt_2^a=\theta^a_1+i\vt^a_1=0$) components
of each $N=8$ superfield are independent
$N=4$ superfields. Moreover, these $N=4$ superfields prove to be subjected to the additional constraints
(which also follow from \p{sumultiplet})
\be\label{sucon}
D^aD_a v=\bD_a \bD^a v=0\,,\quad D^a\varphi=0\,,\; \bD_a \bar\varphi=0\,.
\ee
The $N=4$ superfields $\varphi,\bar\varphi$ comprise the standard $N=4, d=1$ chiral multiplet $({\bf 2,4,2})$,
while the $N=4$ superfield $v$ subject to \p{sucon} and having the $({\bf 1, 4, 3})$ off-shell content
is recognized as the one employed in \cite{{leva2},{ikp}} for constructing the $N=4$ superconformal mechanics
and $N=4$ supersymmetric quantum mechanics with partially broken supersymmetry.

An immediate question is as to how the previously introduced constant $m$ reappears in this new setting.
The answer can be found in \cite{leva2}. First of all, from the constraints \p{sucon} it follows that
\be\label{mm1}
\frac{\partial}{\partial t} \left[ D^a, \bD_a \right] v = 0 \;\Rightarrow \left[ D^a, \bD_a \right] v=\mbox{const}.
\ee
Secondly, rewriting the constraint \p{constm} in terms of the new variables we obtain
\be\label{mm2}
\left.  \left(\nabla^\alpha_i \nabla_{j\alpha}V^{ij}+D^a_i D_{ja} V^{ij}\right) \right|= -3 \left[ D^a, \bD_a \right] v=
6m \Rightarrow  \left[ D^a, \bD_a \right] v=-2m .
\ee
Thus, the constant $m$ now is recovered as a component of the $N=4$ superfield $v$. As we already know from
\cite{leva2}, the presence of this constant in $v$ is crucial for generating the scalar potential term in the action.

The realization of the implicit $N=4$ supersymmetry on the $N=4$ superfields
$v,\varphi,\bar\varphi$ is as follows:
\be\label{n4su}
\delta^* v=\eta_a D^a \bar\varphi+\bar\eta{}^a\bD_a \varphi\,,\quad \delta^*\varphi=-\eta_a D^a v\,,\quad
\delta^*\bar\varphi =-\bar\eta{}^a\bD_a v\,.
\ee
It is worth noting that in the considered $N=4$ formulation, as distinct from the previous one, the constant $m$ does not
appear explicitly in the transformations of the implicit $N=4$ supersymmetry (cf. \p{n4transf}). It is hidden inside the
superfield $v$ and shows up only at the level of the component action (see Sec. 4.1).

To summarize, we observed a new interesting phenomenon: four different off-shell $N=4$ supermultiplets with
the component contents $({\bf 3,4,1})$, $({\bf 0,4,4})$, $({\bf 1,4,3})$ and $({\bf 2,4,4})$ can be paired in two
different ways to yield the same $N=8$ tensor supermultiplet $({\bf 3, 8, 5})$.
In a forthcoming paper \cite{BIKL} we show that in fact all $N=8, d=1$ supermultiplets admitting
$OSp(4^\star|4)$ as a superconformal group
have two different $N=4$ `faces', where either $OSp(4^\star|2)$ or $SU(1,1|2)$
$N=4$ superconformal symmetries are manifest. In Sect. 5 we demonstrate that the  $N=8$ supermultiplet with the
off-shell content $({\bf 5,8,3})$ employed in \cite{DE}, besides the known splitting
$({\bf 3,4,1})\oplus({\bf 2,4,2})$ in
terms of $N=4$ `tensor' and chiral supermultiplets, also admits a representation in terms of the $N=4$ superfield
$v$ obeying \p{sucon} and the $N=4, d=1$ `hypermultiplet'. This yields an alternative $({\bf 1,4,3})\oplus({\bf 4,4,0})$
splitting of this $N=8$ multiplet.

Now we have all the necessary ingredients to construct invariant actions for the $N=8, d=1$ tensor multiplet
in $N=4$ superspace.

\setcounter{equation}0
\section{N=4 superfield actions for $V^{ij}$}

\subsection{Actions for the (1,4,3)$\oplus$(2,4,2) splitting}
In order to construct an invariant action for the $N=8$ tensor multiplet in $N=4$ superspace,
one can proceed from any of its two alternative $N=4$ superfield representations described above.
We shall consider first the actions in terms of the $N=4$ scalar superfield $v$ subjected to the constraints
\p{sucon} and the chiral superfields $\varphi,\bar\varphi$, i.e. for the splitting described in Sect. 3.2.
The equivalent action in terms of the $N=4$ tensor multiplet $v^{ij}$
and the fermionic variant of the hypermultiplet $\xi^i$, eq. \p{n4b}, is naturally constructed in
the $N=4, d=1$ harmonic superspace of ref. \cite{IL}. It will be considered in the next subsection.

The $N=8$ supersymmetric free action for the supermultiplet $v, \varphi, \bar\varphi$ reads
\be\label{free}
S^{free}=-\frac{1}{4} \int dt d^4 \theta \left( v^2 -2 \varphi\bar\varphi \right).
\ee
It is easy to check that it is indeed invariant with respect to the hidden $N=4$ Poincar\'e supersymmetry \p{n4su}
which completes the manifest $N=4$ one to $N=8$.

The free action \p{free} is not invariant with respect to (super)conformal transformations. In order to
construct a $N=8$ superconformal invariant action for $N=4$ superfields $v, \varphi, \bar\varphi$, we follow
a strategy which goes back to \cite{hi} and was applied in \cite{IKL} for constructing
a $N=4$ superconformal action for the $N=4$ tensor
multiplet in $N=2, d=1$ superspace.

The `step-by-step' construction of \cite{IKL} adapted to the given case works as follows. Let us start
from the $N=4$ superconformal invariant action for the superfield $v$ \cite{leva2}
\footnote{In \p{step1} and the subsequent $N=4$ actions we always
assume that the $N=4$ superfields containing the dilaton (i.e. $v$ in the present case) start with a constant,
hence no actual singularities appear.}
\be\label{step1}
S_0=-\frac{1}{4}\int dt d^4\theta \;L_0\,, \quad L_0 =  v \log v\,.
\ee
The variation of $L_0$ with respect to \p{n4su}, up to a total derivative, reads
\be\label{step2}
\delta L_0=-\frac{\delta(\varphi\bar\varphi)}{v} \;.
\ee
Thus, in order to ensure $N=8$ supersymmetry, we have to add to $L_0$ in \p{step1} a term which cancels \p{step2}, namely
\be\label{step3}
L_1 = \frac{\varphi\bar\varphi}{v}.
\ee
Continuing this recursive procedure further, we find the full Lagrangian yielding the $N=8$ supersymmetric action
\bea\label{step4}
L=&& v\left[ \log v +\sum_{n=1}^\infty (-1)^{n}\frac{(2(n-1))!}{n!n!}
\left( \frac{\varphi\bar\varphi}{v^2}\right)^n \right]\nn
&& = v\left[ \log \left( v+\sqrt{v^2+4\varphi\bar\varphi}\right) +1-\log 2\right]-\sqrt{v^2+4\varphi\bar\varphi}\;.
\eea
Discarding terms which do not contribute to the action as a consequence of the basic constraints \p{sucon},
we eventually obtain
\be\label{action}
S=-\frac{1}{4}\int dt d^4\theta \left[ v\log \left( v+\sqrt{v^2
+4\varphi\bar\varphi}\right) -\sqrt{v^2+4\varphi\bar\varphi}\;\right].
\ee

The last step is to check that the action \p{action} is invariant with respect to $N=4$ superconformal transformations
and so is $N=8$ superconformal (since the closure of $N=4$ superconformal and $N=8$ Poincar\'e supersymmetries
is $N=8$ superconformal symmetry).

The subgroup of \p{n8s}- \p{Vtr} with respect to which the $N=4$ superspace \p{IIsup} is closed
(i.e. $SU(1,1|2)$) is singled out
by the following identification of the transformation parameters
\be\label{sctr1}
\epsilon_{1a}=-i\varepsilon_{1a} \equiv \frac{\epsilon_a}{\sqrt{2}},\;
\epsilon_{2a}=i\varepsilon_{2a}=-\frac{\beps_a}{\sqrt{2}}\;.
\ee
The corresponding subset of $N=8$ superconformal transformations is formed just by those transformations
which leave intact the
combinations $\theta^a_2-i\vt_2^a$ and $\theta^a_1+i\vt^a_1$ anticommuting with both $D^a$ and $\bD_a\,$.
The $N=4$ superconformal transformations of the superfields $v, \varphi, \bar\varphi$ have the following form:
\be\label{scrt2}
\delta v=-2i\left( \epsilon_a\bt{}^a+\beps{}^a\theta_a\right)v\,,\quad \delta\varphi=-4i\epsilon_a\bt{}^a\varphi\,,\quad
\delta\bar\varphi=-4i\beps{}^a\theta_a\bar\varphi\,,
\ee
while the full integration measure transforms as
\be\label{scrt3}
\delta \left( dtd^4\theta\right) =2i\left( \epsilon_a\bt{}^a+\beps{}^a\theta_a\right)\left(dtd^4\theta\right)\;.
\ee
The combination $\varphi\bar\varphi/v^2$ is invariant under \p{scrt2}. Thus, only the first
multiplier in \p{step4}, i.e. the superfield $v$, is transformed. Therefore, the transformation of $L$ \p{step4} is
\be\label{scrt4}
\delta L =-2i\left( \epsilon_a\bt{}^a+\beps{}^a\theta_a\right)L
\ee
and cancels out the transformation of the measure \p{scrt3}, ensuring the $N=4$ superconformal invariance of
the action \p{action}. Hence the latter, being also invariant under $N=8$ Poincar\'e supersymmetry, is invariant under
the entire $N=8$ superconformal group $OSp(4^\star|4)$.

The full component action is rather lengthy, but its pure bosonic core is remarkably
simple. Integrating in \p{action} over Grassmann variables, discarding all fermions and
eliminating the auxiliary fields $D^{(a}\bD{}^{b)}v|$ and $(\bD_a\bD{}^a)\varphi$ by their equations of motion,
we end up with the bosonic action\footnote{When passing to the component action we should take
into account that the $N=4$
superfield $v$ contains the constant $m$ among its components, $ \left[ D^a , \bD_a\right] v =-2m$.}:
\be\label{bosaction1}
S_B=\int dt \frac{1}{\sqrt{v^2+4\varphi\bar\varphi}}\left[ {\dot v}{}^2+4\dot\varphi \dot{\bar\varphi}-m^2-
2im{\dot v}-\frac{4im\varphi\dot{\bar\varphi}}{v+\sqrt{v^2+\varphi\bar\varphi}}\right]\;,
\ee
where the fields $v,\varphi,\bar\varphi$ are the first, $\theta$-independent terms in the $N=4$
superfields $v,\varphi,\bar\varphi$, respectively.

The meaning of the action \p{bosaction1} can be clarified by passing to the new field variables
$q,\lambda,\blam $ defined as
\be
v=e^q\left( \frac{1-\lambda\blam}{1+\lambda\blam}\right),\quad \varphi=\frac{e^q\lambda}{1+\lambda\blam},\quad
\bar\varphi=\frac{e^q\blam}{1+\lambda\blam}\;.
\ee
In terms of these fields the action \p{bosaction1} becomes
\be\label{bosaction}
S_B=\int dt\left[e^q {\dot q}{}^2 -m^2 e^{-q} + 4e^q \frac{\dot\lambda\dot\blam}{(1+\lambda\blam)^2}-
2im\frac{\lambda\dot\blam -\dot\lambda\blam}{1+\lambda\blam}\right]\;.
\ee
This amounts to the sum of two conformally invariant $d=1$ actions, i.e. that of conformal mechanics \cite{dff}
and that of a charged particle moving in the field of a Dirac monopole. Thus, the bosonic part of
the action \p{step4} exactly coincides with the bosonic part of the action of $N=4$, $D(2,1;\alpha)$
superconformal mechanics \cite{IKL} corresponding to the particular choice $\alpha=1$, for which
$D(2,1;\alpha)$ becomes $OSp(4^\star|2)$. The mass parameter $m$, which is present among the components
of the original $N=8$ superfield $V^{ik}$, plays the role of the coupling constant, in analogy
with \cite{leva2}. Note that in \cite{IKL} an analogous coupling constant appeared as a free parameter
of the action, while in the considered case it arises as a consequence of the irreducibility
constraints in $N=8, d=1$ superspace.

We conclude that the action \p{action} describes a $N=8$ supersymmetric extension of the conformally
invariant action of the coupled system for the dilaton field and a charged particle in a
Dirac monopole background. The fact that the bosonic sector of the action of this $N=8$ superconformal
mechanics coincides with that of the $\alpha = 1$ case of the $N=4$ superconformal mechanics action associated
with the multiplet $({\bf 3, 4, 1})$ finds its natural explanation in the existence of the alternative
splitting $({\bf 3, 8, 5}) = ({\bf 3, 4, 1})\oplus({\bf 0,4,4})$, whose $N=4$ actions are presented in in
the next Subsection. The second multiplet contains no physical bosons and so does not contribute to the bosonic
sector. The conformal $N=4$ supergroup $OSp(4^\star|2)$ is hidden in the action \p{action}, but it becomes a
manifest symmetry of the alternative $N=4$ action.

\subsection{Superconformal action for the (3,4,1)$\oplus$(0,4,4) splitting}
The action for the splitting of Subsect. 3.1 consists of two pieces. The first one is written in the customary
$N=4, d=1$ superspace and the second one in the analytic subspace of $N=4, d=1$ harmonic superspace \cite{IL}.
So, we first need to rewrite the defining relations of the $N=4$ superfields $v^{(ik)}$ and $\xi^i_\alpha$
(first two eqs. in \p{n4b}) in the harmonic superspace.

We use the definitions and conventions of ref. \cite{IL}. The harmonic variables parametrizing  the coset $SU(2)_R/U(1)_R$
are defined by the relations
\be\label{61}
u^{+i}u^-_i = 1 \quad \Leftrightarrow\quad  u^+_i u^-_j - u^+_j u^-_i = \epsilon_{ij} \;, \;\; \overline{(u^{+i})} = u^-_i\,.
\ee
The harmonic projections of $v^{ik}, \xi^i_\alpha$ are defined by
\bea
&& v^{++}=v^{ij}u^+_i u^+_j\,, \; \xi^+=\xi^i u^+_i,\; \bxi^+=\bxi^i u^+_i\,, \label{62}  \\
&& v^{+-}=\frac{1}{2} D^{--} v^{++}\,, \quad v^{--} = D^{--} v^{+-} \label{63}
\eea
and the relevant part of constraints \p{n4b} is rewritten as
\bea\label{64}
&&D^+ v^{++}=\bD{}^{+} v^{++}=0\;, \quad D^{++}v^{++}=0\;, \nn
&&D^+ \xi^{+}=\bD{}^{+} \xi^{+}=0\;, \; D^{++}\xi^{+}= D^{++}\bxi{}^{+}=0\;.
\eea
Here $D^+ =D^i u^+_i,\; \bD^+ = \bD^i u^+_i\,$,  $D^{\pm\pm} = u^{\pm i}\partial/\partial u^{\mp i}$ (in the central
basis of the harmonic superspace), $D^i, \bar D^i $ are given in \p{defD1} and we use the notation
$(\xi^i, \bar\xi^i) \equiv \xi^{\alpha i}$. The relations \p{64} imply that $v^{++}$ and
$\xi^+, \bar\xi^+$ are analytic harmonic $N=4, d=1$ superfields
living on the invariant subspace $(\zeta, u^\pm_i) \equiv (t_A, \theta^+, \bar\theta^+, u^\pm_i)$.
In this setting, the transformations of the hidden $N=4$ supersymmetry \p{n4transf} are rewritten as
\bea\label{65}
&& \delta^*\xi^+=D^+\bD{}^+\left( 2\bar\eta\,{}^- v^{+-}-\bar\eta{}^+ v^{--}\right) +6m\bar\eta{}^+\,,\nn
&&\delta^*\bxi{}^+=D^+\bD{}^+\left( 2\eta^- v^{+-}-\eta^+ v^{--}\right) +6m\eta^+,\nn
&& \delta^*v^{++}=\eta^+\xi^+-\bar\eta^+\bxi^+ \; , \nn
&& v_{ij}\delta^*v^{ij}=(\eta^+v^{--}-\eta^-v^{+-})\xi^+ -(\eta^+v^{+-}+\eta^-v^{++})\xi^- \nn
&& \quad\quad\quad -(\bar\eta{}^+v^{--} -\bar\eta{}^- v^{+-})\bxi{}^+
+ (\bar\eta{}^+v^{+-} -\bar\eta{}^-v^{++})\bxi{}^-\,.
\eea

The action is given by
\bea\label{66}
S&=&\int dt d^4\theta \sqrt{ v^2}
-\frac{1}{\sqrt{2}}\int du d\zeta^{--}\left[
\frac{\xi^+\bxi{}^+}{\left( 1+c^{--}{\hat v}{}^{++}\right)^{\frac{3}{2}}} \right. \nn
&&\left. +12m
\frac{{\hat v}{}^{++}}{\sqrt{1+c^{--}{\hat v}{}^{++}}(1+\sqrt{1+c^{--}{\hat v}{}^{++}})}\right],
\eea
where $du d\zeta^{--}= du dt_Ad\theta^+d\bar\theta^+ $ is the measure of integration over the analytic superspace and
\be\label{67}
v^{++}={\hat v}{}^{++}+c^{++} \;, \quad c^{\pm\pm} = c^{ik}u^\pm_iu^\pm_k\,, \;\; c^{ik} = \mbox{const}\,.
\ee
The first term in \p{66}  is the superconformal (i.e. $OSp(4^\star|2)$) invariant kinetic term of $v^{ik}$. The second and third ones are
the kinetic term of $\xi^+$ and the superconformal invariant potential term of $v^{ik}$ \cite{IL}.
The action \p{66} is manifestly $N=4$ supersymmetric since it is written in terms of $N=4$ superfields.
However, its invariance with respect to the hidden $N=4$ supersymmetry \p{65} must be explicitly checked. It is enough
to demonstrate the invariance of \p{66} with respect to the $\eta$ part of \p{65}.

The variations of the three terms in \p{66} are
\bea\label{var}
\delta S_1 &=& \int dt d^4\theta \frac{v^{ij}\delta^*v_{ij}}{\sqrt{v^2}} \;, \nn
\delta S_2 &=& \frac{1}{\sqrt{2}}\int du d\zeta^{--} \left[{ { D^+\bD{}^+\left[ \left(2\eta^- v^{+-}-\eta^+ v^{--}\right)
\xi^+\right]}\over{\left(\sqrt{1+c^{--}{\hat v}{}^{++}}\right)^3}} +{{6m\eta^+\xi^+}\over{\left(\sqrt{1+c^{--}{\hat v}{}^{++}}\right)^3}}
\right],  \nn
\delta S_3 &=&-\frac{1}{\sqrt{2}}
\int du d\zeta^{--} \left[{6m\eta^+\xi^+}\over{\left(\sqrt{1+c^{--}{\hat v}{}^{++}}\right)^3}\right].
\eea
It is seen that the $m$-dependent terms are cancelled among $\delta S_2$ and $\delta S_3$. The first term
in $\delta S_2$ may be rewritten as an integral over the full $N=4$ superspace using
$$
\int dtdu d^4\theta = \int dud\zeta^{--}D^+\bar D^+\,.
$$
Thus, the variation of the action \p{66} reads
\be\label{var1}
\delta S=\int dt d^4\theta\left[ \frac{v^{ij}\delta^*v_{ij}}{\sqrt{v^2}} +\frac{1}{\sqrt{2}}
\int du { {  \left(2\eta^- v^{+-}-\eta^+ v^{--}\right)\xi^+ }
\over{\left(\sqrt{1+c^{--}{\hat v}{}^{++}}\right)^3} }   \right].
\ee
Now, using \p{65}, we identically rewrite the numerator in the second term in \p{var1} as
\bea\label{var2}
\left(2\eta^- v^{+-}-\eta^+ v^{--}\right)\xi^+ &=& -v^{ij}\delta^*v_{ij} +\left[ \left( \eta^-v^{+-}\xi^+-\eta^{+}v^{+-}\xi^{-}\right)
+\eta^-v^{++}\xi^-\right]\nn
&&\equiv -v^{ij}\delta^*v_{ij} +\left[ (B_1)+B_2\right].
\eea
The first term in \p{var2} does not depend on harmonics, so the corresponding part of the harmonic integral in \p{var1}
can be easily computed to yield \footnote{This harmonic integral has been
computed in \cite{hi}.}
\be\label{var3}
-\frac{1}{\sqrt{2}}v^{ij}\delta^*v_{ij} \int du \frac{1}{\left(\sqrt{1+c^{--}{\hat v}{}^{++}}\right)^3} =
 -\frac{v^{ij}\delta^*v_{ij}}{\sqrt{v^2}} \;.
\ee
This term is cancelled by the first one in \p{var1}. Finally, we are left with
\be\label{var4}
\delta S = \frac{1}{\sqrt{2}}\int dt d^4\theta du \frac{B_1+B_2}{\left(\sqrt{1+c^{--}{\hat v}{}^{++}}\right)^3}\;.
\ee

Let us prove that the expression
\be\label{var5}
\left(B_1 + B_2\right)\left(1 + X\right)^{-{3\over 2}}\,, \quad X \equiv c^{--}{\hat v}{}^{++}
\ee
is reduced to a total harmonic derivative, modulo purely analytic terms which disappear
under the Berezin integral over the full harmonic superspace.

The proof for the terms $\sim B_1$ and $\sim B_2$ can be carried out independently, so we start with the first term.

As a first step, let us note that $v^{+-}$ in $B_1 =\eta^-v^{+-}\xi^+ - \eta^+v^{+-}\xi^- $ can be replaced by
$\hat{v}^{+-} = v^{+-} - c^{+-}, \;c^{+-} = c^{ik}u^+_iu^-_k\,$. For the first term in $B_1$ this is evident, since
the difference is an analytic superfield. For the second term this follows from the identity
\be\label{proof1}
(\eta^+c^{+-}\xi^-)\left(1 + X\right)^{-{3\over 2}} =  D^{++}\left[
\frac{\eta^+c^{--}\xi^-}{\sqrt{1+X}\left(1 + \sqrt{1+X}\right)}\right],
\ee
which can be easily checked.

Next, we need to show the existence of a function $\tilde{F}^{++}$, such that
\be\label{proof2}
{\hat B}_1 \left(1 +X\right)^{-{3\over 2}} =\left( \eta^-{\hat v}{}^{+-}\xi^+ - \eta^+{\hat v}{}^{+-}\xi^-\right)
\left(1 +X\right)^{-{3\over 2}}=
D^{--}\tilde{F}^{++}\,.
\ee
Choosing for $\tilde{F}^{++}$ the Ansatz
\be
\tilde{F}^{++} = f_1(X) \,(\eta^-\hat{v}^{++}\xi^+) + f_2(X)\,(\eta^+\hat{v}^{++}\xi^-) \,,
\ee
and comparing both sides of \p{proof2}, we find
\be
f_1 = -f_2 = \frac{1}{\sqrt{1+X}\left(1 + \sqrt{1+X}\right)}\,. \lb{f1}
\ee

Proceeding in a similar way, it is straightforward to show that the term proportional to $B_2$
in \p{var5} is also reduced to a total harmonic derivative
\be\label{proof3}
B_2\left(1 + X \right)^{-{3\over 2}} = D^{++}{G}^{--} + \mbox{analytic terms}\,.
\ee
Indeed, making the Ansatz
\be
{G}^{--} = g_1(X)\,(\eta^+c^{--}\xi^-) + g_2(X)\,(\eta^-c^{+-}\xi^-)\,,
\ee
it is easy to find
\be
g_2 = \frac{1}{\sqrt{1 + X}}\,, \quad g_1 = -\frac{1}{1 + \sqrt{1 +X}}\,.
\ee
In obtaining this result, we made use of the relation
\be
c^{++}c^{--} - (c^{+-})^2 = 1\,. \lb{tozhd}
\ee

Thus, we proved that the term \p{var4} vanishes and so the action \p{66} possesses
$N=8$ supersymmetry.

Finally, we should prove the $N=4$ superconformal invariance of the entire action \p{66}. The invariance
of the first and third terms
was shown in \cite{IKL} and \cite{IL}, so it remains for us to prove this property for the second term.

As shown in \cite{IL},
the measure of integration over the analytic superspace is invariant under the $N=4$ superconformal group
($OSp(4^\star|2)$ in our case), while the involved superfields are transformed as
\be\label{68}
\delta \xi^+=\Lambda \xi^+,\; \delta \bxi{}^+=\Lambda \bxi{}^+,\quad
\delta{\hat v}{}^{++}=2\Lambda\left( {\hat v}{}^{++}+c^{++}\right) -2\Lambda^{++}c^{+-}
\ee
where $\Lambda^{++}=D^{++}\Lambda, \; (D^{++})^2\Lambda =0\,$. Here the analytic superfunction $\Lambda$ contains
all parameters of
superconformal transformations (see \cite{IL} for details). The variation of the second term in \p{66} under
these transformations
is
\be
\xi^+\bar\xi^+ \left[ 2\,\frac{\Lambda}{(1+X)^{{3\over 2}}} -3\,\frac{\Lambda\,
(X + c^{++}c^{--})}{(1+X)^{{5\over 2}}} +
3\, \frac{D^{++}\Lambda\, c^{+-}c^{--}}{(1+X)^{{5\over 2}}}\right]. \lb{SCvar1}
\ee
The last term in \p{SCvar1} is vanishing under the harmonic integral, because
$$
c^{+-}c^{--}\,(1 +X)^{-{5\over 2}} = {1\over 2} D^{++}\left[\frac{3 + 3X +  X^2}{(1+X)^{{3\over 2}}
\left[ 1 + (1+X)^{{3\over 2}}\right]}\,(c^{--})^2\right]
$$
and one can integrate by parts with respect to $D^{++}$ and make use of the relations $(D^{++})^2\Lambda = 0$
and $D^{++}\xi^{+} = 0$. Then \p{SCvar1} is reduced to
\be
-\xi^+\bar\xi^+\,\Lambda \left[\frac{1 + X + 3(c^{+-})^2}{(1+X)^{{5\over 2}}}\right]. \lb{SCvar2}
\ee
It can be checked that the expression in the square brackets is equal to
$$
(D^{++})^2 \left[ f_1(X)(c^{--})^2 \right],
$$
where the function $f_1(X)$ already appeared in \p{f1}. Integrating by parts with respect to $(D^{++})^2$ we find
that the variation \p{SCvar2} is vanishing under the harmonic integral, once again due to harmonic constraints
on $\xi^+$ and $\Lambda$.

We finish with two comments.

First, we wish to point out that the actions \p{action} and \p{66}, though describing the same system, are defined on $N=4$
superspaces with drastically different superconformal properties and so cannot be related by any equivalence transformation.
Indeed, the manifest superconformal group of \p{action} realized on the coordinates of the corresponding $N=4$
superspace is $SU(1,1|2)$, and it is the only $N=4, d=1$ superconformal group compatible with chirality. On the contrary,
the manifest superconformal group of \p{66} acting in the relevant $N=4$ superspaces is $OSp(4^\star|2)$.
Both actions have
hidden $N=4$ superconformal symmetries which close on $OSp(4^\star|4)$ together with the manifest ones; they are given
by $OSp(4^\star|2)$
in the first case and $SU(1,1|2)$ in the second one. These hidden superconformal symmetries are not realized on the $N=4$
superspace coordinates, but rather they transform the involved $N=4$ superfields through each other. Thus, the actions
\p{action} and \p{66} bring
to light different symmetry aspects of the same $N=8$ superconformal system. Of course,
these two different $N=4$ actions yield the same
component action; also, they both can presumably be obtained from the single superconformal action of
$V^{ij}$ in $N=8, d=1$ superspace,
e.g. in $N=8, d=1$ harmonic superspace which is a straightforward reduction of the
$N=2, d=4$ one \cite{{harm},{book}}. We
do not address the latter issue in this paper.

The second comment concerns an interesting dual role of the parameter $m$. In the action \p{action}
it is hidden inside the superfield $v$, does not appear
in the transformations of the implicit $N=4$ Poincar\'e supersymmetry \p{n4su} and is revealed only after passing
to the component action, where it produces a scalar potential and magnetic monopole terms
(it yet reappears, due to \p{mm2}, in the anticommutators of the implicit Poincar\'e supersymmetry
and the manifest $SU(1,1|2)$
superconformal symmetry). On the contrary, in the alternative splitting considered here
this parameter appears already in the transformations
of the second $N=4$ Poincar\'e supersymmetry \p{n4transf}, \p{65} and is explicitly present
in the superfield action \p{66} as the coefficient before the
superconformal superfield potential term. As follows from \p{n4transf} and \p{65}, at $m\neq 0$ the
implicit half of $N{=}8, d=1$ Poincar\'e supersymmetry is spontaneously broken, $\xi^i, \bar \xi^i$ being the
relevant Goldstone fermions with an inhomogeneous transformation law. Since, by construction,
all fermions in the theory are the Goldstone ones for the conformal supersymmetry and so transform inhomogeneously
under the latter, we conclude that there is a linear combination of the implicit $N{=}4$ Poincar\'e supersymmetry
generators and half of those of the $N{=}8$ conformal supersymmetry which remains unbroken at $m\neq 0$. At $m=0$ unbroken
is the entire $N{=}8, d{=}1$ Poincar\'e supersymmetry. So the structure of the vacuum symmetries of the model
essentially depends on the parameter $m$.

\setcounter{equation}0
\section{N=8, d=1 vector multiplet}
In this section we shall analyse, along the same line, the off-shell $N=8$ multiplet $({\bf 5, 8, 3})$
considered in \cite{{DE},{BMZ},{Smi1}}. As new results, we shall study its superconformal
properties (which has not been done before), show the existence of a second $N=4$ splitting for it and
construct the corresponding $N=8$ superconformal invariant off-shell actions.

\subsection{Superconformal properties in N=8 superspace}

In analogy with the $N=8$ tensor multiplet, the $N=8$ multiplet $({\bf 5,8, 3})$ employed in \cite{DE}
can also be obtained from the proper nonlinear realization of the supergroup $OSp(4^\star|4)$. The
corresponding supercoset contains the bosonic coset $USp(4)/Spin(4) \sim SO(5)/SO(4)$, four out of five physical
bosonic fields of the multiplet in question being the corresponding coset parameters. The remaining field,
as in the previous case,
is the dilaton associated with the generator $D$. As opposed to the previous case in which the supercoset
contained the
coset $SU(2)_R/U(1)_R$ of the R-symmetry group $SU(2)_R$ (with generators $T^{ik}$), in the present case
this $SU(2)_R$ as a whole
is placed into the stability subgroup. Thus we are going to construct a nonlinear realization
of the superconformal group $OSp(4^\star |4)$
in the coset superspace $\frac{OSp(4^\star|4)}{SU(2)_R \otimes SO(4)}$ parametrized as
\be\label{5coset}
g=e^{itP}e^{\theta_{ia} Q^{ia}+\vt_{i\alpha} \cQ^{i\alpha}}e^{\psi_{ia} S^{ia}+\xi_{i\alpha} \cS^{i\alpha}}
e^{izK}e^{iuD}e^{i v_{\alpha a} U^{\alpha a} }~.
\ee

In order to find superconformal covariant irreducibility conditions on the coset superfields,
we must impose, once again, the inverse Higgs constraints \cite{IH} on the left-covariant $osp(4^\star |4)$-valued
Cartan one-form $\Omega\,$. In full analogy with \p{ih}, we impose the following constraints
\be\label{5ih}
\omega_D=0\; , \quad \left.  \omega^{\alpha a}_U\right| = 0 \; .
\ee
These constraints are manifestly covariant under the left action of the whole supergroup.
They allow one to trade the Goldstone spinor superfields and the superfield $z$ for the spinor
and $t$-derivatives of the remaining bosonic Goldstone superfields $u, v_{\alpha a}$.
Simultaneously, they imply the irreducibility constraints for the latter
\be\label{5constr}
D^{ib}{\cal V}_{\alpha a} + \delta_a^b \nabla_\alpha^i{\cal U}=0\;, \quad
\nabla^{i\beta}{\cal V}_{\alpha a} + \delta_\alpha^\beta D^i_a {\cal U}=0\;,
\ee
where
\be \label{5Lambda}
{\cal V}_{\alpha a}=e^{-u}\frac{2V_{\alpha a}}{2+V^2}\,,\quad {\cal U}=e^{-u} \left( \frac{2-V^2}{2+V^2}\right)\,,\quad
V_{\alpha a}= \frac{\tan \sqrt{\frac{v^2}{2}}}{\sqrt{\frac{v^2}{2}}} v_{\alpha a}\,.
\ee

Note that the superfields $\cU,\cV^{\alpha a}$ provide an example of how to construct a linear representation of $SO(5)$
symmetry in terms of its nonlinear realization. Indeed, with respect to
$SO(4) \sim SU(2)\times SU(2)$ the superfield $\cV^{\alpha a}$ is a 4-vector, in accordance with
its spinor indices,  and  $\cal U$ is a scalar. With respect to the $SO(5)/SO(4)$ transformations generated by
the left action of the group element
\be\label{so5}
g_1=e^{ia_{\alpha a} U^{\alpha a}}\,,
\ee
these superfields transform as
\be\label{so5a}
\delta \cV_{\alpha a}= a_{\alpha a}\; \cU \;, \quad \delta \cU=-2 a_{\alpha a}\; \cV^{\alpha a}\,,
\ee
thus constituting a vector representation of $SO(5)$. Roughly speaking, the $SO(5)/SO(4)$ superfield $V_{\alpha a}$
represents the angular part of this $SO(5)$ vector, while $e^{-u}$ is its radial part. Indeed,
\be
\cU^2 + 2 \cV^2 = e^{-2u}\,.
\ee
Under \p{so5} the coordinates also transform, as
\be\label{so5b}
\delta \theta_{ia}=a_{\alpha a}\vartheta^\alpha_i\;, \quad \delta\vartheta_{i\alpha}=a_{\alpha a}\theta^a_i\;.
\ee

The transformation properties of the superfields $u, V^{\alpha a}$ under the conformal supersymmetry
generated by the left shifts with
\be\label{5g}
g_2=e^{\epsilon_{ia} S^{ia}+\varepsilon_{i\alpha} \cS^{i\alpha}}
\ee
can be easily found
\be\label{5n8s}
\delta u = -2i\left( \epsilon^{ia}\theta_{ia}+\varepsilon^{i\alpha} \vt_{i\alpha}\right)\; , \quad
\delta V_{\alpha a}=2i\left(\delta_\alpha^\beta \delta_a^b +V_\alpha^b V_a^\beta\right) A_{\beta b}
+2iA_{ab}V^b_\alpha+
2i A_{\alpha\beta}V^\beta_a \,.
\ee
Here
\be\label{A}
A_{\alpha a}=\theta_{i a}\varepsilon^i_\alpha+\vt_{i\alpha}\epsilon_a^i\;,\; A_{ab}
=\theta_{ia}\epsilon^i_b+\theta_{ib}\epsilon^i_a\;,\;
A_{\alpha\beta}=\vt_{i\alpha}\varepsilon^i_\beta+\vt_{i\beta}\varepsilon^i_\alpha \;.
\ee

\subsection{Two N=4 superfield formulations}

As in the case of the $N=8$ tensor multiplet, we may consider two different splittings of the $N=8$ vector
multiplet into irreducible $N=4$ superfields. They correspond to two different choices of $N=4, d=1$ superspace
as a subspace of $N=8, d=1$ superspace.

The first $N=4$ superspace is parametrized by the coordinates $\left\{ t, \theta_{ia} \right\}$. It follows from the
constraints \p{5constr} that the spinor derivatives of all involved superfields with respect to $\vartheta_{i\alpha}$
are expressed in terms of spinor derivatives with respect to $\theta_{i a}$. This means that the only
essential $N=4$ superfield components
of $\cV_{\alpha a}$ and $\cU$ in their $\vartheta$-expansion are the first ones
\be\label{n4comp}
\hat{\cV}_{\alpha a} \equiv \cV_{\alpha a}|_{\vt=0}\;,\quad \hat{\cU} \equiv \cU|_{\vt=0}\,.
\ee
They accommodate the whole off-shell component content of the $N=8$ vector multiplet. These five bosonic
$N=4$ superfield are
subjected, in virtue of \p{5constr}, to the irreducibility constraints in $N=4$ superspace
\be\label{5constra}
D^{i(a}\hat{\cV}{}^{b)\alpha}=0,\quad D^{i(a}D_i^{b)} \hat{\cU}= 0.
\ee
Thus, from the $N=4$ superspace perspective, the vector $N=8$ supermultiplet amounts to the sum of the
$N=4,d=1$ hypermultiplet $\hat{\cV}_{\alpha a}$ with
the $({\bf 4,4,0})$ off-shell component content and the $N=4$ `old' tensor multiplet $\hat{\cU}$
with the $({\bf 1,4,3})$ content.

The transformations of the implicit $N=4$ Poincar\'e supersymmetry, completing the manifest
one to the full $N=8$ supersymmetry, have the following form in terms of $N=4$ superfields:
\be\label{5n4transfa}
\delta^*\hat{\cV}_{a\alpha}= \eta_{i\alpha} D^i_a \hat{\cU}\;, \quad \delta^*\hat{\cU}
={1\over 2}\eta_{i\alpha}D^{ia}\hat{\cV}_a^\alpha\;.
\ee

Another interesting $N=4$ superfield splitting of the $N=8$ vector multiplet can be achieved by passing to the
complex parametrization of the $N=8$ superspace as
\be
\left\{ t,\Theta_{ia}= \theta_{ia}+i\vartheta_{ia},
\bar\Theta^{ia}= \theta^{ia} -i\vartheta^{ia}\right\} \lb{3sup},
\ee
with a manifest diagonal $su(2)$ (this is just the parametrization corresponding to the direct reduction
from $N=2, d=4$ superspace, with the spinor derivatives satisfying the algebra \p{T2}). In this superspace
we define
the covariant derivatives $\cD^{i\alpha},\cbD{}^{j\beta}$ as
\be
\cD^{i\alpha}\equiv \frac{1}{\sqrt{2}}\left( D^{i\alpha}- i\nabla^{i\alpha}\right)\;, \quad
\cbD{}^{i\alpha}\equiv \frac{1}{\sqrt{2}}\left(  D^{i\alpha}+ i\nabla^{i\alpha}\right)\;,\quad
\left\{ \cD^{i\alpha}, {\overline\cD}{}^{j\beta}\right\}=2i\epsilon^{ij}\epsilon^{\alpha\beta}\partial_t
\ee
(the indices $\beta $ and $a$ are indistinguishable with respect to the diagonal $SU(2)$)
and pass to the new notation
\bea
\cV \equiv -\epsilon_{\alpha a}\cV^{\alpha a}\,, \quad \cW^{\alpha\beta} \equiv \cV^{(\alpha \beta)} = {1\over 2}
\left( \cV^{\alpha \beta}+\cV^{\beta \alpha}\right)\,, \quad \cW \equiv \cV +i{\cal U}\,,
\quad \overline{\cW} \equiv \cV - i{\cal U}\,.
\eea
In this basis of $N=8$ superspace, the original constraints \p{5constr} amount to
\footnote{This form of constraints is a direct $d=1$
reduction of the $N=2$ superfield constraints of $N=2, d=4$ supersymmetric abelian gauge theory.
They can be solved through the $d=1$ analytic harmonic
gauge potential \cite{BMZ} like their $d=4$ counterparts \cite{{harm},{book}}.}
\bea\label{5constrb}
&&\cD^{i\alpha}\cW^{\beta\gamma}= -\frac{1}{4}\left( \epsilon^{\beta\alpha}\cbD{}^{i\gamma}\overline{\cW}+
\epsilon^{\gamma\alpha}\cbD{}^{i\beta}\overline{\cW}\right),\;
\cbD{}^{i\alpha}\cW^{\beta\gamma}=-\frac{1}{4}\left( \epsilon^{\beta\alpha}
\cD^{i\gamma} \cW+\epsilon^{\gamma\alpha}\cD{}^{i\beta}\cW\right),\nn
&& \cD^{i\alpha}\overline{\cW}=0,\;\cbD^{i\alpha} \cW=0\,,
\quad (\cD^{k\alpha}\cD^i_\alpha) \cW =(\cbD^{k}_{\alpha}\cbD^{i\alpha})\overline{\cW}\,.
\eea
Next, we single out the $N=4, d=1$ superspace in \p{3sup} as
$\left( t, \theta_a \equiv \Theta_{1a}, \bar\theta^{a}\right)$
and split our $N=8$ superfields into $N=4$ ones in the standard way.
As in the previous cases,  the spinor derivatives of each $N=8$ superfield with respect to
$\overline{\Theta}^{2a}$ and $\Theta_{2a}$, due to the constraints \p{5constrb}, are expressed as derivatives
of other superfields with respect to $\bar\theta^{a}$ and $\theta_{a}$. Therefore, only the
first (i.e. taken at $\overline{\Theta}^{2a}=0$ and $\Theta_{2a}=0$) $N=4$ superfield components
of $N=8$ superfields are independent.
They accommodate the entire off-shell field content of the multiplet. These $N=4$ superfields are
defined as
\be
\left. \Phi \equiv \cW \right|\,, \quad, \left. \bar\Phi \equiv \overline{\cW}\right|\,,
\quad \left. W^{\alpha\beta} \equiv \cW^{\alpha\beta}\right|
\ee
and satisfy the constraints following from \p{5constrb}
\be\label{5finalconstr}
{\cD}^{\alpha} \bar\Phi =0\,,\quad \cbD_{\alpha}\Phi=0\,,\quad
\cD^{(\alpha}W^{\beta\gamma)}=\cbD{}^{(\alpha}W^{\beta\gamma)}=0\,, \quad \cD^\alpha
\equiv \cD^{1\alpha}\,, \;\cbD_\alpha \equiv \cbD_{1\alpha}\,.
\ee
They tell us that the $N=4$ superfields $\Phi$ and $\bar\Phi$ form the standard $N=4$ chiral multiplet $({\bf 2, 4, 2})$,
while the $N=4$ superfield $W^{\alpha\beta}$ is recognized as the $N=4$ tensor multiplet $({\bf 3,4,1})$.
Just this $N=4$ superfield set has been proposed in \cite{DE} to represent the vector $N=8, d=1$ supermultiplet
in question. As we see, it provides another `face' of the same $N=8$ multiplet defined by the
$N=8$ superspace constraints \p{5constr}. The first `face' (unknown before) is the set
$({\bf 1, 4, 3})\oplus({\bf 4,4,0})$ considered earlier.

The implicit $N=4$ supersymmetry is realized on $W^{\alpha\beta}$, $\Phi$ and $\bar\Phi$ as
\be\label{6tra}
\delta^*W^{\alpha\beta}=\frac{1}{2}\left(\eta^{(\alpha}\cbD^{\beta)}\bar\Phi
-\bar\eta{}^{(\alpha}\cD{}^{\beta)}\Phi\right)\,,\quad \delta^*\Phi=
\frac{4}{3}\eta_\alpha \cbD^\beta W_\beta^\alpha\,,\quad
\delta^* \bar\Phi=-\frac{4}{3}\bar\eta{}^{\alpha}\cD_\beta W_\alpha^\beta\,.
\ee

\subsection{N=8 supersymmetric actions in N=4 superspace}

As we showed, the $N=8$ vector multiplet has two different descriptions in terms of $N=4$ superfields,
the `new' and `old' ones.
The first (new) description involves the $N=4$ `old tensor' multiplet $\hat{\cU}$
and the `hypermultiplet' $\hat{\cV}_{a \alpha}$ obeying the constraints \p{5constra}.

For constructing the $N=4$ superfield superconformal action for this case, we start with the following Ansatz:
\be\label{6actionb1}
L=L_0+\sum_{n=1}A_n= L_0 + \frac{1}{\cU}\sum_{n=1} a_n\left( \frac{\hat{\cV}{}^2}{\hat{\cU}{}^2}\right)^n\,,
\quad L_0 = \frac{1}{\hat{\cU}}\,,
\ee
where $L_0$ is the $OSp(4^\star|2)$ superconformal invariant action for $\cU$ \cite{IKL2}.

The complete action \p{6actionb1}, by construction, possesses $N=4$ superconformal $OSp(4^\star|2)$ invariance.
Indeed, with respect to the
particular subset of transformations \p{5n8s} with
\be\label{dok1}
\epsilon_{1a}=\epsilon_a\,,\quad \epsilon_{2a}=-\beps_a\,,\quad \varepsilon_{i\alpha}=0 \;,
\ee
which do not mix the superfields $\hat{\cU}$ and $\hat{\cV}_{a\alpha}$ and preserve
the $N=4$ superspace $(t, \theta_{ia}) \equiv
(t, \theta_a, -\bar\theta_a)\,$, the measure transforms as
\be\label{dok2}
\delta (dt d^4\theta) = 2i(\epsilon_a \bt{}^a+\beps{}^a\theta_a)(dt d^4\theta)\;,
\ee
while
\be\label{dok3}
\delta (V^2) =0
\ee
and, therefore,
\be\label{dok4}
\delta (\hat{\cV}^2) =4i(\epsilon_a \bt{}^a+\beps{}^a\theta_a)(\hat{\cV})^2,\quad \delta \hat{\cU}=
2i(\epsilon_a \bt{}^a+\beps{}^a\theta_a)\hat{\cU}
\ee
(recall the relations \p{5Lambda}). Since $\left(\hat{\cV}{}^2/ \hat{\cU}{}^2\right)^n$
is invariant with respect to \p{dok4},
it follows from the representation \p{6actionb1} that $L$ transforms as
\be\label{dok5}
\delta L = -2i(\epsilon_a \bt{}^a+\beps{}^a\theta_a)L\,,
\ee
which cancels the variation of the measure \p{dok2}.

Now we should choose the coefficients $a_n$ so as to make the action \p{6actionb1}
invariant with respect to the implicit
$N=4$ supersymmetry \p{5n4transfa}. The variation of the first term under \p{5n4transfa}
can be represented as
\be\label{6actionb2}
\delta L_0 = \frac{\delta (\hat{\cV}{}^2)}{2\,\hat{\cU}{}^3} \quad \Rightarrow a_1=-{1\over 2} \;.
\ee
Going further, one finds the following recurrence relation between the coefficients $a_n$ which ensures
the invariance of the action \p{6actionb1}:
\be\label{6actionb3}
a_{n+1}=-\frac{(2n+1)}{(n+2)(n+1)}a_n\quad \Rightarrow a_n=\frac{(-1)^n (2n-1)!}{2^{n-1}(n+1)!(n-1)!}\;.
\ee

Let us note that the same relations among the coefficients $a_n$ \p{6actionb3} appear if we require
the invariance of the action \p{6actionb1} under the $SO(5)/SO(4)$ transformations
\p{so5a}, \p{so5b} which, in terms of $N{=}4$ superfields, look as follows
\be\label{dop1}
\delta^*{\hat\cV}_{\alpha a}=a_{\alpha a} {\hat\cU}+a_{\alpha b} \theta^b_i D^i_a {\hat\cU}\,,\quad
\delta^*{\hat\cU}=-2a_{\alpha a} {\hat\cV}{}^{\alpha a}+\frac{1}{2} a_{\alpha a} \theta^a_i D^{ib}
{\hat\cV}{}^\alpha_b\;.
\ee
Indeed, the variation of $L_0$ after integrating by parts can be cast in the form
\be\label{so1}
\delta L_0 =\frac{1}{\hat{\cU}{}^2}\left(X+\frac{Y}{\hat\cU}\right), \mbox{  where  }
X\equiv a_{\alpha a}{\hat\cV}{}^{\alpha a},\;
Y\equiv a_{\alpha a}\theta^a_i D^i_b {\hat\cU} {\hat\cV}{}^{\alpha b},
\ee
while the variation of generic term $A_n$ in \p{6actionb1} is as follows:
\be\label{so2}
\delta A_n=2a_n\left[ n+ \frac{(2n+1)(n+1)}{n+2} \frac{{\hat\cV}{}^2}{{\hat\cU}^2}\right]
\frac{{\hat\cV}{}^{2(n-1)}}{{\hat\cU}^{2n}}
\left( X+ \frac{Y}{\hat\cU}\right).
\ee
It is easy to see that, with $a_n$ as in \p{6actionb3}, all terms in the $SO(5)$ variation of \p{6actionb1}
are cancelled among themselves.

Summing up all terms, we get the $N=8$ supersymmetric Lagrangian in the closed form
\be\label{6actionb4}
L= \frac{2}{{\hat\cU} +\sqrt{{\hat{\cU}{}^2}+2\hat{\cV}{}^2}}.
\ee
Thus, the action
\be\label{dok5a}
S= 2\int dt d^4\theta\,\frac{1}{{\hat\cU} +\sqrt{{\hat{\cU}{}^2}+2\hat{\cV}{}^2}}
\ee
possesses both $N=4$ superconformal symmetry and $N=8$ supersymmetry. The closure of these
two supersymmetries yields the whole $N=8$ conformal supergroup $OSp(4^\star |4)$, which
thus represents the full invariance group of \p{dok5a}.

The second possibility includes the $N=4$ tensor $W^{\alpha\beta}$ and the chiral
$\Phi$, $\bar\Phi$ multiplets \p{5finalconstr}.

The free action invariant under the implicit $N=4$ supersymmetry \p{6tra} reads
\be\label{6actiona1}
S_{free}=\int dt d^4\theta \left( W^2-\frac{3}{4}\Phi\bar\Phi\right).
\ee
The $N=4$ superconformal invariant action which is invariant also under \p{6tra} has the very simple form
\be\label{6actiona2}
S_{kin}=2\int dt d^4 \theta\;{{\log \left( \sqrt{W^2}+\sqrt{W^2+\frac{1}{2}\Phi\bar\Phi}\right)}\over\sqrt{W^2}}\,.
\ee
The proof of $N=4$ superconformal symmetry in this case (it is now $SU(1,1|2)$) is similar
to that performed in Section 4.1.
Indeed, under the restriction \p{sctr1} the transformation of the measure is the same as in \p{scrt3}
and
\be\label{dok6}
\delta (W^2) =4i(\epsilon_a \bt{}^a+\beps{}^a \theta_a ) (W^2) \,,
\; \delta \bar\Phi=4i\epsilon_a \bt{}^a \bar\Phi\,,\;
\delta \Phi=4i\beps{}^a \theta_a \Phi\,.
\ee
Once again, one may be easily convinced that the variation of the Lagrangian in \p{6actiona2} cancels the variation
of the measure. This proves the $N=4$ (and hence $N=8$) superconformal invariance of the action \p{6actiona2}.

The bosonic part of the action \p{6actiona2} has the very simple form
\be\label{final1}
S_{B}=\int dt \; \frac{{\dot W}_{\alpha\beta}{\dot W}{}^{\alpha\beta} +\frac{1}{2} \dot\Phi \dot{\bar\Phi} }
{ \left( W^2 +\frac{1}{2}\Phi\bar\Phi \right)^{\frac{3}{2}}}\,.
\ee
It is invariant under the $SO(5)$ transformations \p{so5} which are realized on the bosonic fields as
\be\label{final2}
\delta \Phi=-i a \Phi -2i a_{\alpha\beta} W^{\alpha\beta},\;
\delta\bar\Phi = i a \bar\Phi +2i a_{\alpha\beta} W^{\alpha\beta},\quad
\delta W_{\alpha\beta}=-\frac{i}{2} a_{\alpha\beta}\left( \Phi-\bar\Phi \right),
\ee
where
\be\label{final3}
a \equiv -\epsilon_{\alpha a}a^{\alpha a}\,, \quad a^{\alpha\beta} \equiv a^{(\alpha \beta)} = {1\over 2}
\left( a^{\alpha \beta}+a^{\beta \alpha}\right).
\ee
This action supplies a particular case of the $SO(5)$ invariant bosonic target metric found in \cite{DE}.
The most general $SO(5)$ invariant metric of \cite{DE} can be reproduced from the superfield
action constructed as the sum of \p{6actiona2} and the non-conformal (but $SO(5)$ invariant) action \p{6actiona1}.

Let us recall that the $N=4, d=1$ tensor multiplet admits a $N=4$ superconformal invariant potential term \cite{IKL,IL},
but the latter can be written only in terms
of the prepotential for $W^{\alpha\beta}$, or as an integral over the analytic harmonic superspace \cite{IL}.
This term can be promoted to a $N=8$ superconformally invariant one.
However, this issue is out of the scope of the present paper. We plan to consider it elsewhere.

\setcounter{equation}0
\section{Conclusions}
In this paper we constructed new models of $N{=}8$ superconformal mechanics associated with the
off-shell multiplets $({\bf 3, 8, 5})$ and $({\bf 5, 8, 3})$ of $N{=}8, d{=}1$ Poincar\'e supersymmetry.
We showed that both multiplets can be described in a $N{=}8$ superfield form
as properly constrained Goldstone superfields associated with suitable cosets of the
nonlinearly realized $N{=}8, d{=}1$
superconformal group $OSp(4^\star|4)$. The $N{=}8$ superfield irreducibility conditions
were derived as a subset of superconformal covariant constraints on the Cartan super one-forms.
The superconformal transformation properties of these $N{=}8, d{=}1$ Goldstone superfields were
explicitly given, alongside with the transformation of the coordinates of $N{=}8, d{=}1$ superspace.
Although these superfield irreducibility constraints can also be reproduced, via dimensional reduction,
from the analogous constraints defining off-shell $d{=}4$ tensor and vector multiplets in $d{=}4$ superspace,
our method, being self-contained in $d{=}1$, is not tied to such a procedure. Also, the field contents of
the considered $d{=}1$ multiplets differ essentially from those of their $d{=}4$ ancestors.

Apart from the $N{=}8$ superfield description, we presented several $N{=}4$ superfield formulations
of these $d{=}1$ multiplets.
As an interesting phenomenon, we revealed the existence of two different $N{=}4$ `faces' of each single $N{=}8$ multiplet.
These alternative $N{=}4$ formulations
deal with different pairs of off-shell $N{=}4$ multiplets (see \p{1} and \p{2}) and generically manifest
different $N{=}4$ superconformal subgroups of $OSp(4^\star|4)$. More concretely for both (\ref{1}) and (\ref{2}),
the first splitting features the superconformal $N{=}4$ subgroup $SU(1,1|2)$ (with some central charges added)
while the second one involves $OSp(4^\star|2)$.
We constructed $N{=}8$ superconformal invariant off-shell $N{=}4$ superfield actions for all four splittings considered.
To our knowledge, these actions have not yet appeared in the literature.
As a subsymmetry of $OSp(4^\star|4)$ they all respect a hidden $USp(4) \sim SO(5)$ symmetry.
Hence, for the multiplet $({\bf 5,8,3})$ our superconformal actions provide a special case of the $SO(5)$ invariant
$N{=}4$ superfield action of \cite{DE} (which presented the target metric but not the action).

An obvious project for future study is to investigate the implications of our $N{=}8$ superconformal mechanics models
for the AdS$_2$/CFT$_1$ correspondence. Indeed, the $N{=}4$ superconformal mechanics associated with
the multiplet $({\bf 3, 4, 1})$ \cite{IL} is believed to be equivalent, both classically and
quantum-mechanically \cite{{ikn},{bgik}},
to a superparticle on the supercoset $\frac{SU(1,1|2)}{SO(1,1)\otimes U(1)}$ which is a superextension of
AdS$_2\times S^2$. One of the $N{=}8$ models considered here, namely the one associated with
the multiplet $({\bf 3, 8, 5})$, is
a further superextension of the $N{=}4$ superconformal mechanics just mentioned,
with the bosonic sector unchanged. It is therefore
natural to expect a relation to some AdS$_2\times S^2$ superparticles defined
on appropriate cosets of the supergroup $OSp(4^\star|4)$ since it contains both $SU(1,1|2)$ and
$OSp(4^\star|2)$ as supersubgroups. The superconformal $({\bf 5, 8, 3})$ SQM model can
also have an interesting application in a similar context. As argued in \cite{GTW}, this type of $N{=}8$ superconformal
mechanics may be relevant to the near-horizon AdS$_2\times S^4$ geometry of a D5-brane in an orthogonal
D3-brane background. The supergroup $OSp(4^\star|4)$ just defines the corresponding superisometry.

As proposed in \cite{Smi1}, the models associated with the multiplet $({\bf 5, 8, 3})$ can describe the
dimensionally reduced Coulomb branch
of $d{=}4$ Seiberg-Witten theory, and the corresponding actions may be connected with the low-energy quantum
effective action of this theory.
It would be interesting to elucidate the role of the superconformal invariant model as well as
the existence of its two different but equivalent $N{=}4$ descriptions from this point of view.

Our main focus in this paper was on super{\it conformally\/} invariant actions of the $N{=}8$ multiplets
considered. One could equally well
employ these off-shell multiplets for constructing $N{=}8$ supersymmetric, but not superconformal,
$d{=}1$ models. It is desirable to learn
what is the most general bosonic target geometry associated with such sigma models.

Finally, let us recall that, besides $OSp(4^\star|4)$, there exist other $N{=}8, d{=}1$ superconformal groups, namely
$OSp(8|2)\,$, $F(4)$ and $SU(1,1|4)$ (see e.g. \cite{VP}). It is clearly of interest to set up nonlinear realizations
of all these supergroups, deducing the corresponding
$N{=}8$ multiplets and constructing the associated $N{=}8$ supersymmetric and superconformal mechanics models.
Certain steps towards this goal will be presented in a forthcoming paper \cite{BIKL}.

\section*{Acknowledgments}
S.K. would like to thank B.~Zupnik for many useful discussions.
This work was partially supported by the European Community's Human Potential
Programme under contract HPRN-CT-2000-00131 Quantum Spacetime,
the INTAS-00-00254 grant, the NATO Collaborative Linkage Grant PST.CLG.979389,
RFBR-DFG grant No 02-02-04002, grant DFG No 436 RUS 113/669, RFBR grant
No 03-02-17440 and a grant of the Heisenberg-Landau programme.
E.I. and S.K. thank the Institute for Theoretical Physics of the University of Hannover
and the INFN--Laboratori Nazionali di Frascati
for the warm hospitality extended to them during the course of this work.

\newpage

\def\theequation{A.\arabic{equation}}
\section*{Appendix: Superalgebra $osp(4^\star |4)$}

Here we present the explicit structure of the superalgebra $osp(4^\star|4)$ in the basis \p{alg8}.

Boson-boson sector:
\bea\label{bb}
&& i\left[ D, P\right]=P,\quad i\left[D, K\right]=-K,\quad i\left[ P, K\right]=-2 D, \quad
 i\left[ T^{ij},T^{kl}\right]=\epsilon^{ik}T^{jl}+\epsilon^{jl}T^{ik},\nn
&& i\left[ T_1^{ab},T_1^{cd}\right]=\epsilon^{ac}T_1^{bd}+\epsilon^{bd}T_1^{ac},\;
  i\left[ T_2^{\alpha\beta},T_2^{\gamma\rho}\right]=\epsilon^{\alpha\gamma}T_2^{\beta\rho}+
 \epsilon^{\beta\rho}T_2^{\alpha\gamma},\nn
&& i\left[T_1^{ab},U^{c\alpha}\right]=\frac{1}{2}\left( \epsilon^{ac}U^{b\alpha}+\epsilon^{bc}U^{a\alpha}\right),\;
i\left[T_2^{\alpha\beta},U^{a\gamma}\right]=\frac{1}{2}\left( \epsilon^{\alpha\gamma}U^{a\beta}+
 \epsilon^{\beta\gamma}U^{a\alpha}\right),\nn
&&i\left[ U^{a\alpha},U^{b\beta}\right]=
2\left( \epsilon^{\alpha\beta}T_1^{ab}+\epsilon^{ab}T_2^{\alpha\beta}\right)\;.
\eea

Mixed boson-fermion sector:
\bea\label{bf}
&& i\left[P, Q^{ia}\right]=0,\; i\left[P, \cQ^{i\alpha}\right]=0,\;
i\left[P, S^{ia}\right]=-Q^{ia},\; i\left[P, \cS^{i\alpha}\right]=-\cQ^{i\alpha},\nn
&& i\left[D, Q^{ia}\right]=\frac{1}{2}Q^{ia},\; i\left[D, \cQ^{i\alpha}\right]=\frac{1}{2}\cQ^{i\alpha},\;
i\left[D, S^{ia}\right]=-\frac{1}{2}S^{ia},\; i\left[D, \cS^{i\alpha}\right]=-\frac{1}{2}\cS^{i\alpha},\nn
&& i\left[K, Q^{ia}\right]=S^{ia},\; i\left[K, \cQ^{i\alpha}\right]=\cS^{i\alpha},\;
i\left[K, S^{ia}\right]=0,\; i\left[K, \cS^{i\alpha}\right]=0,\nn
&& i\left[ U^{a\alpha},Q^{ib}\right]=\epsilon^{ab}\cQ^{i\alpha},\;
 i\left[ U^{a\alpha},\cQ^{i\beta}\right]=\epsilon^{\alpha\beta}Q^{ia},\;
i\left[ U^{a\alpha},S^{ib}\right]=\epsilon^{ab}\cS^{i\alpha},\nn
&& i\left[ U^{a\alpha},\cS^{i\beta}\right]=\epsilon^{\alpha\beta}S^{ia},\;
 i\left[T^{ab},Q^c\right]=\frac{1}{2}\left( \epsilon^{ac}Q^{b}+\epsilon^{bc}Q^{a}\right).
\eea

Fermion-fermion sector:
\bea\label{ff}
&& \left\{ Q^{ia},Q^{jb}\right\}=-2\epsilon^{ij}\epsilon^{ab}P,\;
 \left\{ \cQ^{i\alpha},\cQ^{j\beta}\right\}=-2\epsilon^{ij}\epsilon^{\alpha\beta}P,\nn
&& \left\{ S^{ia},S^{jb}\right\}=-2\epsilon^{ij}\epsilon^{ab}K,\;
 \left\{ \cS^{i\alpha},\cS^{j\beta}\right\}=-2\epsilon^{ij}\epsilon^{\alpha\beta}K,\nn
&& \left\{ Q^{ia},S^{jb}\right\}=2\left(\epsilon^{ab}T^{ij}-\epsilon^{ij}\epsilon^{ab}D-
     2\epsilon^{ij}T_1^{ab}\right),\nn
&&  \left\{ Q^{ia},\cS^{j\alpha}\right\}=-2\epsilon^{ij}U^{a\alpha},\quad
 \left\{ \cQ^{i\alpha},S^{ja}\right\}=-2\epsilon^{ij}U^{a\alpha},\nn
&& \left\{ \cQ^{i\alpha},\cS^{j\beta}\right\}=2\left(\epsilon^{\alpha\beta}T^{ij}-
  \epsilon^{ij}\epsilon^{\alpha\beta}D-
     2\epsilon^{ij}T_2^{\alpha\beta}\right).
\eea

\end{document}